\documentclass[a4paper,fleqn,usenatbib]{mnras}

\usepackage[T1]{fontenc}
\usepackage{ae,aecompl}

\usepackage{graphicx}	
\usepackage{amsmath}	
\usepackage{amssymb}	

\usepackage{txfonts}

\newcommand{\myurl}[1]{\url{#1}}
\usepackage{hyperref}	
\hypersetup{colorlinks=true,linkcolor=blue,citecolor=blue,filecolor=blue,
urlcolor=blue}

\newcommand{\msol}{\mathrm{M}_\odot}

\newcommand{\mvir}{M_{200,\,\mathrm{cr}}}

\newcommand{\rvir}{r_{200,\,\mathrm{cr}}}

\newcommand{\machnum}{\mathcal{M}}

\newcommand{\mytilde}{\raise.17ex\hbox{$\scriptstyle\sim$}}

\newcommand{\mybf}[1]{#1}

\newcommand{\comment}[1]{}

\title[Hydrodynamic shocks in the Illustris universe]{Shock finding on a 
moving-mesh: II. Hydrodynamic shocks in the Illustris universe}

\author[K.~Schaal et al.]{\parbox{17cm}{Kevin Schaal$^{1,2}$\thanks{e-mail: 
kevin.schaal@h-its.org}, Volker Springel$^{1,2}$, R\"udiger Pakmor$^{1}$, Christoph Pfrommer$^{1}$, 
Dylan Nelson$^3$, Mark Vogelsberger$^4$, Shy Genel$^5$\thanks{Hubble Fellow}, Annalisa Pillepich$^6$, Debora Sijacki$^7$, Lars Hernquist$^6$}
  \vspace*{0.2cm}\\
  $^1$Heidelberg Institute for Theoretical Studies, Schloss-Wolfsbrunnenweg 
35, D-69118 Heidelberg, Germany\\
  $^2$Zentrum f\"ur Astronomie der Universit\"at Heidelberg,
  Astronomisches Recheninstitut, M\"{o}nchhofstr. 12-14, 69120
  Heidelberg, Germany\\
$^3$Max-Planck-Institut
f\"{u}r Astrophysik, Karl-Schwarzschild-Stra\ss{}e 1, 85740 Garching
bei M\"{u}nchen, Germany\\
$^4$Department of Physics, Kavli Institute for Astrophysics \& Space Research, MIT, Cambridge, MA 02139, USA\\
$^5$Department of Astronomy, Columbia University, 550 West 120th Street, New York, NY 10027, USA\\
$^6$Harvard-Smithsonian Center for Astrophysics, 60 Garden Street, Cambridge, MA 02138, USA\\
$^7$Institute of Astronomy and Kavli Institute for Cosmology, University of Cambridge, Madingley Road, Cambridge CB3 0HA, UK}

\date{Published: 4.7.2016}

\pubyear{2016}

\begin{document}
\label{firstpage}
\pagerange{\pageref{firstpage}--\pageref{lastpage}}
\maketitle

\begin{abstract}
  Hydrodynamical shocks are a manifestation of the non-linearity of
  the Euler equations and play a fundamental role in cosmological gas
  dynamics. In this work, we identify and analyse shocks in the
  Illustris simulation, and contrast the results with those of
  non-radiative runs. \mybf{We show that simulations with more
  comprehensive physical models of galaxy
  formation pose new challenges for shock finding algorithms due to
  radiative cooling and star-forming processes, prompting us to
  develop a number of methodology improvements.}
  We find in Illustris a total shock surface
  area which is about 1.4 times larger at the present epoch compared to non-radiative runs,
  and an energy dissipation rate at shocks which is higher by a factor of
  around 7. Remarkably, shocks with Mach numbers above and below
  $\mathcal{M}\approx10$ contribute about equally to the total
  dissipation across cosmic time. \mybf{This is in sharp contrast to non-radiative
  simulations, and we demonstrate that a
  large part of the difference arises due to strong black hole radio-mode
  feedback in Illustris. We also provide an overview of the large diversity of shock morphologies,
  which includes complex networks of halo-internal
  shocks, shocks on to cosmic sheets, feedback shocks due to black
  holes and galactic winds, as well as ubiquitous accretion shocks.}
  In high redshift systems more massive than
  $10^{12}\,\mathrm{M}_\odot$ we discover the existence of a double accretion shock
  pattern in haloes. They are created when gas streams along filaments
  without being shocked at the outer accretion shock, but then forms a
  second, roughly spherical accretion shock further inside.
\end{abstract}


\section{Introduction}
\label{sec:introduction}

Hydrodynamical shock waves play an important role in the evolution of
the baryonic component of our Universe. During the process of
hierarchical structure formation, shocks dissipate and thermalize
kinetic energy in supersonic gas flows, thereby heating the cosmic web
and structures therein.  Gas inside dark matter haloes
can be compressed and virialized by means of shocks, before radiative
cooling in the densest regions allows it to collapse further,
eventually leading to the formation of stars and accretion on to
black holes (BHs).  Supersonic gas motions are also created by feedback
processes such as stellar winds, supernovae (SNe) explosions, or
relativistic jets from active galactic nuclei (AGN). The conversion of
the released mechanical energy to thermal energy in the ambient gas is
again mediated by thermalization processes that ultimately involve
hydrodynamical shocks.

Unlike in classical hydrodynamical shocks in which the kinetic energy
of a fluid is thermalized via particle collisions, shocks on
cosmological and galactic scales are \textit{collisionless}. This is
because the mean free path of particles in the corresponding
environments is large compared to the transition layer between the
pre-shock and post-shock regions. In this case, the energy transfer is
mediated by electromagnetic viscosity \citep{WENTZEL_1974,
  KENNEL_1985}.  Interestingly, collisionless shocks are also sites of
diffusive shock acceleration \citep[DSA;][]{AXFORD_1977,
  KRYMSKI_1977, BELL_1978a, BELL_1978b, BLANDFORD_1978, MALKOV_2001},
creating relativistic cosmic rays. In this process, ions inside the
converging flow can cross the shock multiple times by scattering off
the magnetic field fluctuations in the pre- and post-shock regions, thereby
gaining more and more energy with every crossing. The net result is a
power law momentum spectrum of energetic particles, thus creating a
non-thermal cosmic ray particle component that may modify the gas
dynamics in important ways. This in particular happens in SN
remnants, but may also be important for accretion shocks around
dark matter haloes and AGN.

Directly observing cosmological shocks is in general very difficult.
While the central high density regions of galaxy clusters can be
observed in X-ray, the gas temperature in these environments is high
as well. This allows only for low Mach number shocks, and these are
difficult to detect since the density contrast between the shocked and
unshocked gas is small. Still, a number of bow shocks in merging
clusters have been discovered \mybf{by analysing \textit{XMM-Newton} 
observations \cite[e.g.][]{FINOGUENOV_2010, OGREAN_2013a} and}
\textit{Chandra} observations \cite[e.g.][]{Markevitch2005, Macario2011, 
Russell2014, DASADIA_2016}, the most famous being the Mach number
$\machnum\approx 3$ shock in the bullet cluster
\citep{MARKEVITCH_2002}. 
\mybf{Moreover, pressure jumps associated with shocks
have been detected with the \textit{Planck} space telescope at the outskirts 
of the Coma cluster \citep{PLANCK_COMA, ERLER_2015}.}

Fortunately, the synchrotron emission of accelerated or reaccelerated
electrons can reveal the location of a cosmological shock more easily.
Relativistic cosmic ray electrons have short lifetimes, and their
non-thermal emission is therefore confined to a region close to the
accelerating shock.  There is strong evidence that radio relics
\citep[e.g.][]{FERETTI_2005, FERRARI_2008, BRUEGGEN_2011}, which can
be found in peripheral cluster regions, are associated with merger
shocks \citep{ENSSLIN_1998, MINIATI_2001, PFROMMER_2008a,
PFROMMER_2008b, PINZKE_2013, SKILLMAN_2013}. 
These sources are elongated and often curved, with length-scales 
up to a few $\mathrm{Mpc}$.
Moreover, the emitted photons are highly
polarized and indicate in many cases a magnetic field which is aligned
with the relic \citep{BRUEGGEN_2012}.  Strikingly, for a few clusters
the associated shock is directly observed in the form of an
X-ray edge \citep[e.g.][]{MARKEVITCH_2010, AKAMATSU_2013,
 BOURDIN_2013, OGREAN_2013, OWERS_2014}.

\mybf{On the other hand, for a few merging clusters the presence of X-ray surface 
brightness discontinuities does not correlate with observable radio emission,
or the spectral index of the radio emission implies a Mach number which is inconsistent with
the density jump detected in X-rays \citep[e.g.][]{RUSSELL_2011, OGREAN_2013b, OGREAN_2014}.
These cases may be influenced by projection effects, but they may also 
indicate that DSA directly from the thermal pool is
not sufficient, in particular at weak shocks.}

Radio relics are very interesting objects to study with the current
generation of radio observatories, but they are in principle only the
brightest components of a much larger entity, the magnetized cosmic web.
The cosmic web consists of non-linear structures in the form of a
network of filaments and nodes, which is expected to be surrounded by
more or less stationary accretion shocks.  Based on current estimates
for the magnetisation and electron acceleration efficiencies at these
shocks, the expected synchrotron radiation will be in reach of the
Square Kilometre Array (SKA) \citep{KESHET_2004, 
BATTAGLIA_2009, VAZZA_2015}. 
Consequently, in the near future it should become 
possible to observe filaments of the cosmic web and their surrounding
shocks, providing a novel probe of cosmic large-scale structure
formation and the simultaneous opportunity to test simulation models.

In this paper, we analyse shock waves in an advanced cosmological
simulation of galaxy formation, the Illustris Simulation.  It is the
second work in a short series on shocks in numerical hydrodynamic
simulations. In our first paper \citep{SCHAAL_2015}, we have
introduced our shock finding algorithm which operates on the Voronoi
mesh of the {\small AREPO} code \citep{SPRINGEL_2010}, and analysed
non-radiative cosmological simulations of gas and dark matter.  Those
simulations did not incorporate cooling or feedback of any kind, and
hence the shocks which are present can mostly be classified as merger shocks
and accretion shocks. In this case, the shock statistics features a
bimodality where so-called external shocks enter with high Mach
numbers and internal shocks inside structures with relatively low Mach
numbers \citep{RYU_2003}. The external shocks are created when
pristine gas accretes on to the cosmic web and gets shock heated for
the first time. While the corresponding flows are highly supersonic,
only a little kinetic energy is thermalized due to the low densities involved.
On the other hand, internal flows are characterized by high
temperatures and densities, which lead to very energetic low Mach
number shocks.

Shocks in such non-radiative simulations have been analysed
extensively in previous studies (\mybf{see for example \citet{QUILIS_1998, 
MINIATI_2000, RYU_2003, PFROMMER_2006, SKILLMAN_2008, VAZZA_2009, 
VAZZA_2010, SCHAAL_2015}, and \citet{VAZZA_2011} for a code comparison
project), but only few authors have explored runs including radiative
physics and stellar feedback \citep[e.g][]{KANG_2007, PFROMMER_2007, 
PLANELLES_2013, HONG_2014, HONG_2015} or AGN feedback \citep{VAZZA_2013, VAZZA_2014}.}
It is therefore the aim of our second paper in
this series to investigate shocks in a state-of-the-art cosmological
simulation that includes gas and dark matter, as well as stars and
BHs and their associated feedback processes. These feedback
sources are expected to create powerful additional shocks, making it
particularly interesting to compare the resulting shock statistics
with those obtained for the non-radiative runs.  We shall use the
Illustris simulation suite \citep{VOGELSBERGER_2014,
  VOGELSBERGER_2014b, GENEL_2014} for our study, focusing on the
highest-resolution, full physics run (Illustris-1), also known as
\textit{the} Illustris simulation. This is among the presently most successful 
simulations of galaxy formation
and uses an accurate mesh-based hydrodynamical method, making it
particularly well suited for our purposes.

The paper is structured as follows. Section~\ref{sec:recap} recaps the
analysed simulations, the simulation technique, and the shock finding
method.  Section~\ref{sec:shock_statistics} focuses on the
interpretation of shock statistics, both in a global sense and within 
different environments. In
Section~\ref{sec:morphologies}, we present shock morphologies across
cosmic time and discuss what they can reveal about the gas dynamics.
Variations in our shock finding techniques that test our methodology
are investigated in Section~\ref{sec:methodology_variations}, and our
findings are summarized and discussed in Section~\ref{sec:summary}.
Lastly, we study resolution effects in
Appendix~\ref{sec:resolution_study}.

\section{Analysing shocks in the Illustris universe}
\label{sec:recap}

\subsection{The Illustris project}
\label{sec:illustris_simulations}

At the centre of the Illustris project is the cosmological
hydrodynamical simulation Illustris-1 \citep{VOGELSBERGER_2014,
  VOGELSBERGER_2014b, GENEL_2014}. It evolves around $1.2\times 10^{10}$
resolution elements of dark matter and baryons in a comoving periodic
volume $75\,h^{-1}\,\mathrm{Mpc}$ on a side. With this setup a dark
matter mass resolution of $6.26\times10^6\,\msol$ and an initial gas
mass resolution of $1.26\times10^6\,\msol$ are achieved.  The
simulations of the Illustris project adopt the standard
$\mathrm{\Lambda}$ cold dark matter model and cosmological parameters according to
the 9-year \textit{Wilkinson Microwave Anisotropy Probe}
(\textit{WMAP}9) observations \citep{WMAP9}, with the following
values: $\Omega_{\mathrm{m}}=0.2726$, $\Omega_{\Lambda}=0.7274$,
$\Omega_{\mathrm{b}}=0.0456$, $\sigma_8=0.809$, $n_\mathrm{s}=0.963$,
and $H_0=100\,h\,\mathrm{km\,s^{-1}\,Mpc^{-1}}$ where $h=0.704$.

The simulations of Illustris were carried out with the moving-mesh
code {\small AREPO} \citep{SPRINGEL_2010} and a galaxy formation
physics implementation described and validated in detail in
\citet{VOGELSBERGER_2013} and \citet{TORREY_2014}. 
 In the following, we briefly summarize the
key features of the physics model adopted for Illustris-1. The
interstellar medium (ISM) is represented through a sub-resolution
approach with an effective equation of state following
\citet{SPRINGEL_2003}. In this model, a self-regulated ISM arises through
the interplay of gas cooling, star formation, supernova (SN) feedback
and galactic winds. Star particles are created stochastically in
overdense regions according to the estimated local star formation
rate, and represent single-age stellar populations characterized by a
\citet{CHABRIER_2003} initial mass function (IMF). Additionally,
SNe energy is injected as purely kinetic energy of wind
particles. The latter are briefly decoupled from the hydrodynamic
scheme until they leave the dense star-forming region and can deposit
their momentum into the surrounding lower density gas. Black holes
(BHs) and the associated energy feedback processes from gas accretion
are implemented following \citet{SPRINGEL_2005} and
\citet{SIJACKI_2007}.

\mybf{Collisionless BH sink particles are
inserted at the potential minima of newly identified dark matter haloes
that do not yet contain BHs. To be precise, a BH
particle is seeded if the friends-of-friends group is 
more massive than $5\times10^{10} h^{-1}\msol$, and its initial mass is set to 
$10^{5} h^{-1}\msol$. A repositioning scheme ensures that BH particles
always stay in the minimum of their halo potential.
They grow in mass by accretion and BH merger events. 
The accretion is parametrized by means of a
Bondi-Hoyle-Lyttleton accretion rate
\citep{HOYLE_1939, BONDI_1944, BONDI_1952}, and
limited by the Eddington rate. Theses rates are calculated 
with respect to the parent gas cell of a BH sink particle, 
and in every timestep mass is drained from this cell and transferred to the BH.}

For high accretion rates relative to the Eddington rate, the BH is
assumed to be in a quasar-mode, and the model couples the feedback
energy thermally to the surrounding gas. For low accretion rates, a
different feedback channel, a radio-mode, is assumed in which mechanical
feedback occurs through jets.  Since the jets cannot be directly
resolved, the hot radio bubbles inflated by them are instead modelled
through an injection of heat energy in spherical volumes corresponding
to the bubbles. Additionally, a scheme for radiative AGN feedback is
adopted in which the elevated ionizing flux and its impact on the
cooling rates is accounted for in the proximity of actively accreting
BHs.  The total radiative heating rate of the gas is
calculated by the superposition of the AGN radiation field and a
spatially uniform but time-dependent photoionizing UVB
(UVB). The cooling rate in the simulations has three contributions:
atomic cooling based on the equations describing the ionization
balance of hydrogen and helium \citep{KATZ_1996}, metal line cooling
calculated from pre-computed cooling tables, and Compton cooling off the cosmic
microwave background.  Nine different elements are explicitly tracked
in the simulation (H, He, C, N, O, Ne, Mg, Si, Fe), whose abundance is
increased by stellar winds from AGB stars, core collapse SNe
(SN), and type-Ia SNe.

With this setup of the Illustris simulations, a significant number of
basic properties of the galaxy population beyond those used in tuning
the model (which are mostly the present-day stellar mass function and
the cosmic star formation rate history) have been 
found to be in good agreement with observations. 
Remarkably, these include realistic morphologies and
velocity structures for around 40000 well-resolved galaxies
\citep{VOGELSBERGER_2014,VOGELSBERGER_2014b,
GENEL_2014, SNYDER_2015, TORREY_2015}.
The simulation also contains 32552 BHs at $z=0$ with
properties that match observed quasar luminosity functions and
galaxy-BH correlations quite well \citep{SIJACKI_2015}.  The star
formation histories and stellar masses of galaxies show a plausible
diversity \citep{SPARRE_2015}, and the hydrogen reionization history
of the Universe can be successfully explained by the high-redshift
galaxies forming in Illustris \citep{BAUER_2015}.  All simulation data
has recently been made publicly available \citep{NELSON_2015b}, and a
comprehensive list of further studies analysing the simulation data
can be found online\footnote{\label{foot:illustirs_web_pate}
  See\hspace*{0.15cm} \url{
    http://www.illustris-project.org/results/}}.

While the focus of this work is on the analysis of shocks in the
highest-resolution full physics Illustris-1 simulation, we also show
results for the lower resolution counterparts Illustris-2 and
Illustris-3, and a few selected results for the non-radiative
simulations Illustris-NR-2 and Illustris-NR-3.  Here the numbers 2 and
3 indicate an 8 and 64 times lower mass resolution compared to
Illustris-1, respectively. We note that a more extensive analysis of
the non-radiative simulations can be found in the first paper of the
series \citep{SCHAAL_2015}.

\subsection{Moving-mesh hydrodynamics}
\label{sec:moving_mesh_hydro}

The cosmological simulations analysed in this work were run with the
{\small AREPO} code \citep{SPRINGEL_2010}, which solves the Euler
equations of ideal hydrodynamics on a moving Voronoi mesh by means of
a second order finite volume method. This approach improves on
traditional hydrodynamic solvers such as smoothed-particle
hydrodynamics (SPH) and grid-based schemes by combining advantages of each. 
In particular, the Euler equations are solved accurately
by using a finite volume method on the unstructured Voronoi mesh. The
mesh-generating points are moved with the local velocity field,
resulting in very small residual mass fluxes across cell interfaces
and hence a quasi-Lagrangian and manifestly Galilean-invariant
behaviour. This is desirable in the cosmological structure formation
problem, because it allows the accurate tracking of highly supersonic
flows without large advection errors and additional timestep
constraints. Also, the natural adaptivity of the mesh resolution means
that the mass per resolution element is automatically kept
approximately constant.

Recently, \citet{PAKMOR_2016} introduced refinements for the
moving-mesh scheme in the form of an improved gradient estimator and a
suitable Runge-Kutta time integrator. They have shown that with these
changes {\small AREPO} conserves angular momentum to a high degree and
the convergence rate of the code is improved. We note, however, that
the Illustris-1 simulation still used an older version of the code.

Moving-mesh simulations are well suited for the analysis of
hydrodynamic shocks in cosmology.  In a finite volume scheme, the
solution is discontinuous across cell interfaces and no artificial
viscosity is required, allowing shocks to be captured in principle
over one or very few cells. Furthermore, with the moving-mesh approach
the resolution at shocks tends to be increased because the compression
of the gas also implies a compression of the grid cells transverse to
the shock front, creating an effective enhancement of the resolution
particularly in the post-shock region. This feature is very beneficial
for constructing an accurate shock finding algorithm, as outlined in
the subsequent section.

\begin{figure*}
\centering
\includegraphics{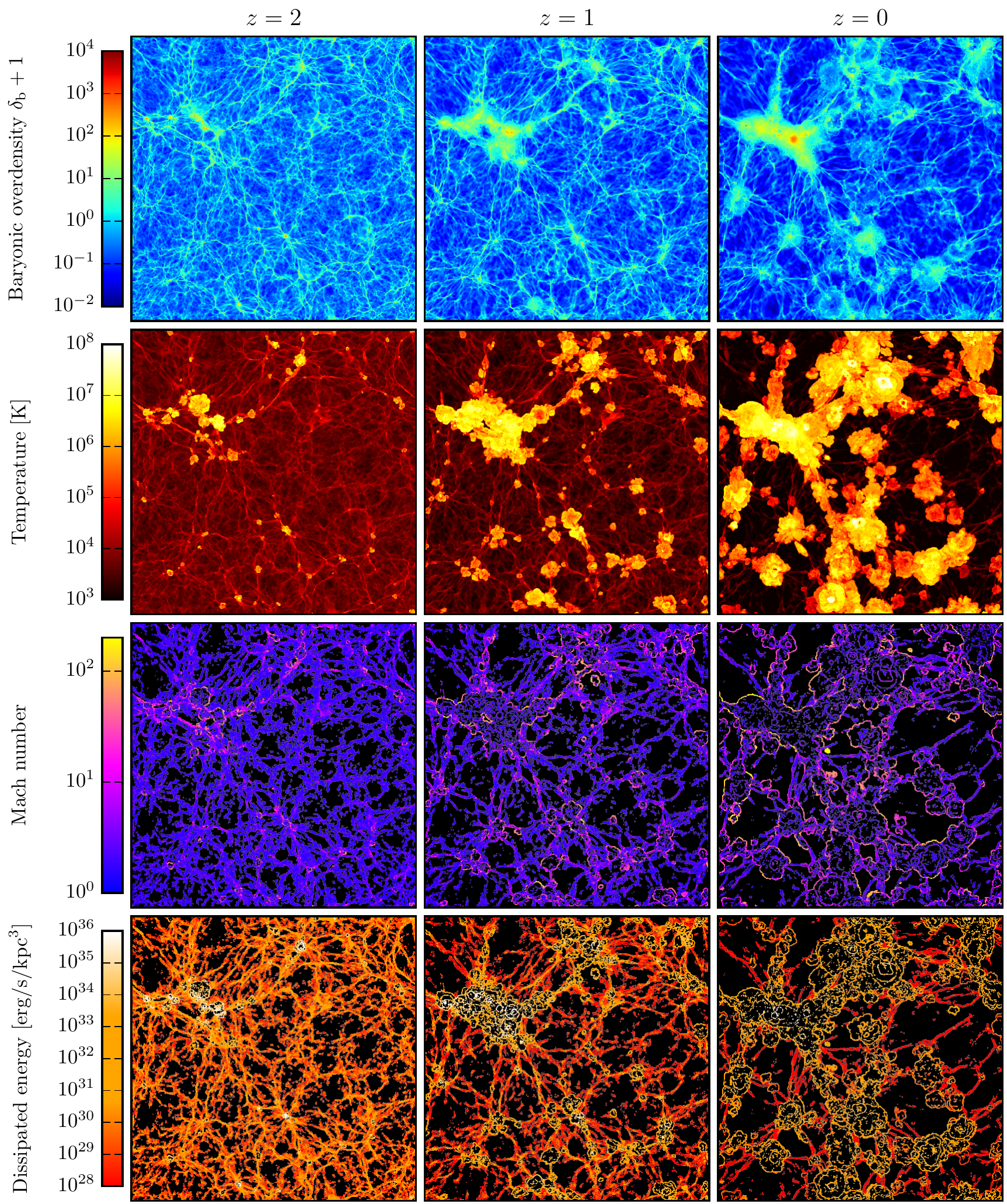}
\caption{Projections of the baryonic overdensity, the temperature, the
  Mach number field, and the energy dissipation rate density for the
  full Illustris-1 box $(75\,h^{-1}\mathrm{Mpc}$ per side) at
  different redshifts.  The $z-$coordinate of the projections is
  centred on one of the most massive clusters in the simulation which
  can be seen in the upper left of each panel
  ($\mvir=2.2\times10^{14}\,\msol$ at $z=0$).  On large scales,
  accretion shocks on to the cosmic web dominate in the early universe,
  whereas at late times the radio mode feedback of BHs is
  omnipresent and creates hot bubbles that shock heat the ambient
  medium. When cold gas from voids gets shocked, Mach numbers of
  several hundreds can be reached, but the overall energy dissipated
  is dominated by the cosmic web.  A movie of the shock dynamics in a
  sub box of Illustris-1 may be accessed online:
  \url{https://youtu.be/1CXCRv3i3uw}.}
\label{fig:fullbox}
\end{figure*}

\subsection{Shock finding}
\label{sec:shock_finding}

Constructing an accurate method for identifying hydrodynamic shocks in
cosmological simulations proves to be challenging. They are
numerically broadened across a few cells and local criteria are
therefore not sufficient. Moreover, tangential discontinuities arising
in shear flows or contact waves are omnipresent and must not be
confused with shocks by the algorithm.

Our method for revealing shocks in moving-mesh simulations is
described comprehensively in \citet{SCHAAL_2015}.  It is based on a
generalization of previous algorithms \citep{RYU_2003, SKILLMAN_2008,
  HONG_2014} to an unstructured mesh and incorporates a number of
additional improvements.  Here we briefly outline the main steps of
the algorithm.  

First, we flag cells in regions around shocks, which we call
\textit{shock zones} \citep[as in][]{SKILLMAN_2008}.  Specifically, a
cell is considered part of a shock zone if and only if all of the
following three criteria are met.  (i) The gas inside the cell is 
compressed (negative velocity divergence), (ii) the temperature and
density gradient must point in the same direction, and (iii) a minimum
Mach number jump ($M\approx 1.3$) is exceeded.

In a second step, the shock is reconstructed as a single layer of
cells containing the \textit{shock surface}. The corresponding cells
are given as those shock zone cells that exhibit the maximum
compression along the shock normal, which is defined by means of the
temperature gradient. Finally, the Mach number is estimated from the
Rankine--Hugoniot temperature jump condition based on the pre- and
post-shock temperature values directly outside of the shock zone.

Once the Mach number of the shock is known, its energy 
dissipation can be calculated, which is the fraction of kinetic energy
\textit{irreversibly} converted into thermal energy. The corresponding
thermal energy flux $f_{\text{th}}$ is given by the product
of the thermalization efficiency $\delta(\machnum)$ \citep{RYU_2003,
  KANG_2007} and the inflowing kinetic energy flux,
\begin{align}
f_\text{th}=\delta(\machnum)\frac{1}{2}\rho_1(c_1\machnum)^3,
\end{align}
where $\rho_1$ and $c_1$ are the pre-shock density and pre-shock sound
speed, respectively.
  
In \citet{SCHAAL_2015}, we have demonstrated that our algorithm can
reliably characterize shocks in non-radiative simulations. However,
the full physics run Illustris-1 incorporates cooling, star formation
and feedback, which poses additional challenges for the shock finding
algorithm. It turns out that this necessitates additional algorithmic
measures in the detection of shocks in full physics runs, and we hence
extend our methodology as follows.

First of all, we need to take care of the star-forming
regions. Because the pressure of cells with active star formation is
set artificially such that the gas follows an effective equation of
state based on a sub-grid multiphase ISM model \citep{SPRINGEL_2003},
the sharp transition in the thermodynamic properties between
star-forming and non star-forming cells can potentially be interpreted
as a shock by our algorithm, distorting our shock statistics.  We
therefore exclude all cells which lie in density-temperature phase
space close to the effective equation of state. Specifically, we
ignore detections for which either the pre-shock cell, the post-shock
cell, or the cell with the shock fulfils $\rho>\rho_\text{sfr}$ and
$T<10^5(\rho/\rho_\text{sfr})^{0.2}\,\mathrm{K}$, where
$\rho_\text{sfr}$ is the star formation threshold. \footnote{\mybf{The excluded
region is indicated by a grey dashed line in the mass-weighted
density-temperature histogram discussed in
Section~\ref{subsub:envzero}.}}

Furthermore, gas cooling and fragmentation due to self-gravity can
sometimes result in large gradients of the primitive variables which
can appear as correspondingly strong jumps when these gradients are
not well-resolved locally. In such a case, the inferred jumps are not
consistent with the Rankine--Hugoniot conditions across shocks and
should hence not be taken into account. Addressing this issue
is primarily important in high density regions. In these
locations strong cooling can be present, and moreover,
measured energy dissipation rates of spurious detections contribute significantly
to the overall dissipation.

We therefore adopt two
further precautions in our shock finding algorithm when applied to
full physics runs.  First, for all tentative shock detections
we compare the pre-shock values of pressure, density, and
velocity to the corresponding post-shock values, and ignore 
a detection if one of these variables change in the wrong direction.
Recall that the shock direction in our algorithm is determined
by means of the temperature gradient.

Secondly, for overdense regions with pre-shock densities
$\delta_\mathrm{b}>1000$, we calculate in addition to the Mach number
based on the temperature jump ($\machnum_T$), the Mach numbers based
on the pressure jump ($\machnum_p$) and density jump
($\machnum_\rho$).  We then check the mutual consistency of the Mach
numbers, and only keep detections with
$f^{-1}\machnum_p<\machnum_T<f\machnum_p$, and
$f^{-1}\machnum_\rho<\machnum_T<f\machnum_\rho$.  The latter filter is
only applied to detections with $M_\rho<3$ or $M_T$ < 3, since the
density jump is not sensitive for high Mach numbers. 
\mybf{If the different Mach numbers of a cell are 
consistent, we keep the Mach number inferred by the temperature jump.}

With the parameter $f$ introduced here, the tolerance for accepting
detections that deviate from the expected jump conditions can be
adjusted. From a theoretical point of view, if a jump in the
hydrodynamic variables represents a shock, the Mach numbers inferred
by the different jump conditions should be perfectly equal.  However,
this equality can of course not be expected when measuring shock
parameters in numerical simulations due to the unavoidable
discretization errors, and consequently, $f$ should be set to a value
well above unity. In this work we adopt $f=2$.

The presence of the unwanted effects described above is demonstrated
in Section~\ref{sec:methodology_variations}, where we also show that
our approach for coping with these limitations is effective. Most
importantly, the adopted numerical parameter of $f=2$ is a robust
choice; varying it within the interval $[1.3,4.0]$ gives very similar
results.  This finding indicates that spurious detections of shocks due
to poorly resolved local cooling and fragmentation effects indeed
strongly violate the Rankine--Hugoniot jump conditions, as otherwise
they could not be filtered out so easily.

We want to note that for strongly radiating gas, the jump conditions
across shocks deviate from the classic Rankine--Hugoniot conditions
derived for non-radiating gas. This is especially the case for
regions where the cooling time-scale is very small. In our
algorithm the densest regions around star-forming gas are not taken
into account, and we hence assume that we can achieve sufficient
accuracy by using the unmodified Rankine--Hugoniot
jump conditions.

When we analyse the non-radiative run Illustris-NR-2, an important
issue arises due to its lack of cosmic reionization.  This leads to
void regions that are unrealistically cold at low redshift, and hence
the strength of shocks in gas that streams out of voids and
accretes on to the cosmic web is overestimated. In order to obtain
more realistic Mach numbers for these shocks, we impose a temperature
floor of $10^4\,\mathrm{K}$ on the gas in the post-processing of
non-radiative simulations, similar as done in previous studies
\citep{RYU_2003, SKILLMAN_2008, SCHAAL_2015}. 
\mybf{This simple approach is sufficient for our purposes, since the 
overall energy dissipation at shocks is not very sensitive 
to the post-processing reionization model. Nevertheless, 
we note that a more realistic model can be obtained by means
of a fitting procedure \citep{VAZZA_2009}.}

\section{Shock statistics}
\label{sec:shock_statistics}

\subsection{Global shock statistics}
\label{sec:global_statistics}

\subsubsection{The global picture}

\begin{figure}
\centering
\includegraphics{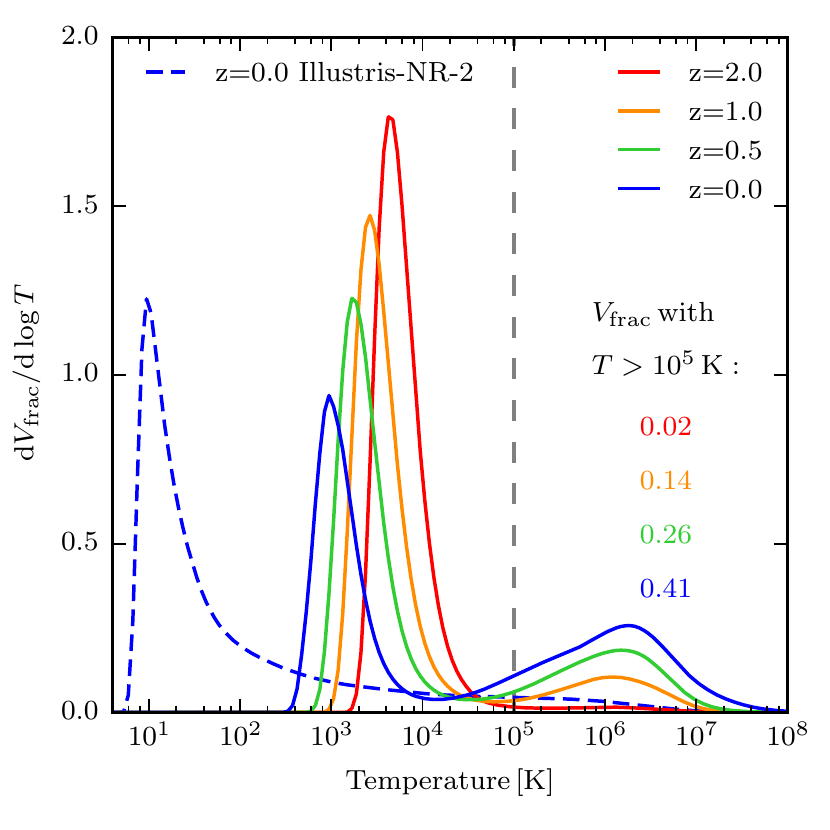}
\caption{Differential volume fraction of the gas in the Illustris-1
  simulation as a function of temperature with respect to the whole
  simulation volume.  Due to BH feedback processes the
  distribution develops a bimodality at late times; at redshift zero a
  significant volume fraction ($\approx 40\%$) is filled with gas
  hotter than $10^5\,\mathrm{K}$.  The positions of the left peaks
  indicate typical void temperatures in Illustris-1 for the different
  redshifts. \mybf{The blue dashed line shows the volume fraction for a 
  non-radiative run as a reference.}}
\label{fig:tfraction}
\end{figure}

\begin{figure*}
\centering
\includegraphics{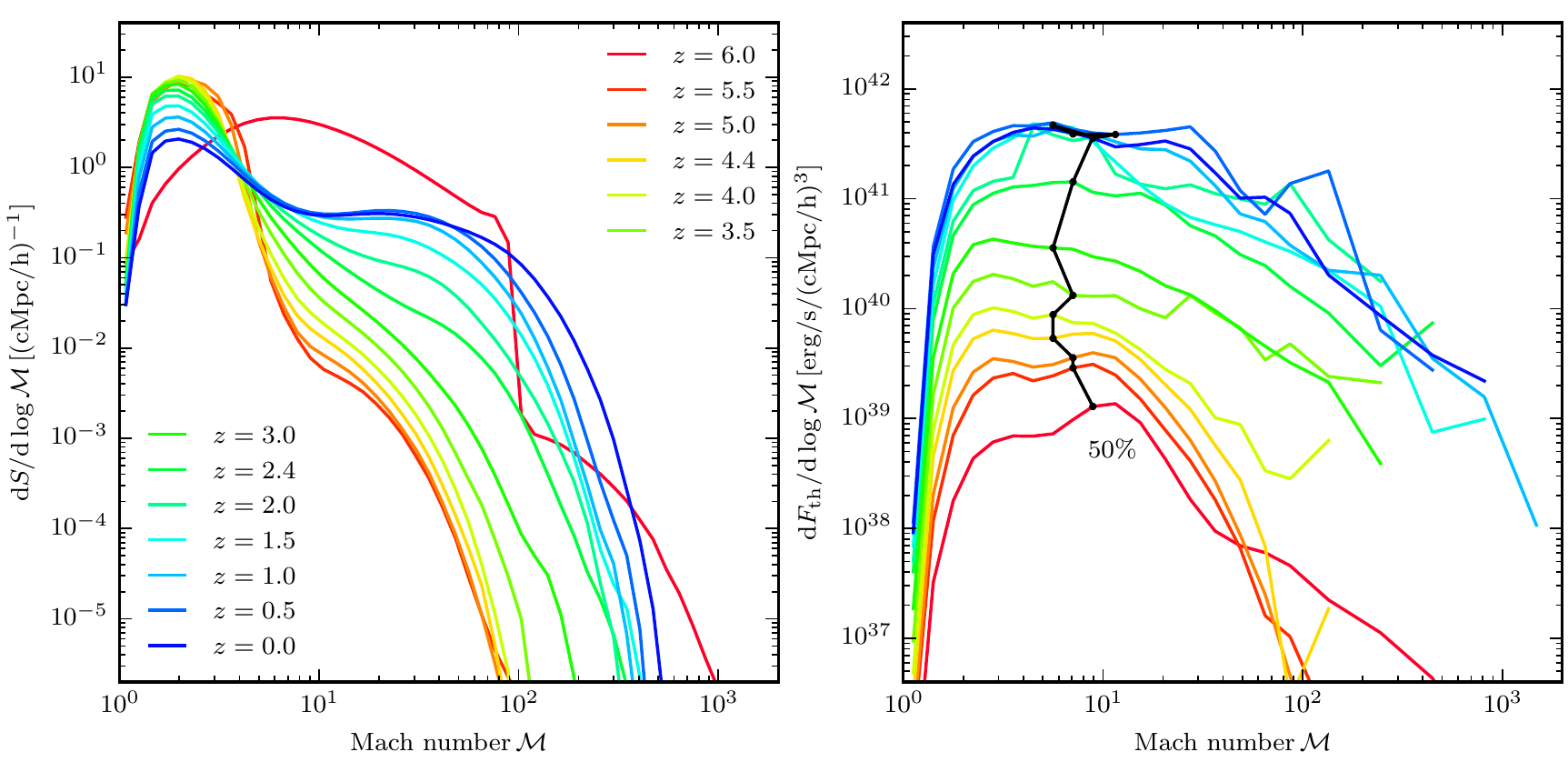}
\caption{{\em Left-hand panel:} shock surface
  area per volume as a function of Mach number; this statistic is
  dominated by external shocks on to the cosmic web. The red line
  indicates the shock distribution on the eve of reionization, 
  during which voids are heated to $10^4\,\mathrm{K}$ 
  and the distribution is shifted towards lower Mach numbers. 
  Subsequently,
  voids merge and cool adiabatically, resulting in a decrease of low
  Mach number shocks and a larger number of shocks with high Mach
  numbers.  {\em Right-hand panel:} energy dissipation at shocks as a
  function of Mach number.  At all redshifts, weak and strong shocks
  contribute similarly to the total thermalization.  This is in
  contrast to the dissipation in non-radiative runs, for which high
  Mach number shocks are less important. Consequently, the increased
  dissipation at the high Mach number end can be attributed to shocks
  created by feedback processes.  }
\label{fig:surface_energy}
\end{figure*}

\mybf{We start by investigating global shock patterns in the
Illustris-1 simulation.} Fig.~\ref{fig:fullbox} presents projections of
the mean baryonic overdensity, the mass-weighted temperature, the
dissipated energy weighted Mach number, and the mean dissipated energy
density at different redshifts.  For the density and temperature, the
projection depths are $100\,\mathrm{kpc}$, whereas for the quantities
inferred from the shock finder we adopt a smaller value of
$50\,\mathrm{kpc}$ in order to better show the thin shock
surfaces. The extent of each panel corresponds to the width of the
full simulation volume of $75\,h^{-1}\mathrm{Mpc}$ and is centred on
one of the most massive objects at $z=0$. The mass inside $\rvir$, the
radius for which the mean overdensity of the halo equals 200 times the
critical density of the universe, is $\mvir=2.2\times10^{14}\,\msol$
for this central object, similar to the most massive cluster in the
volume with $\mvir=2.3\times10^{14}\,\msol$.

At redshift $z=2$, corresponding to a lookback time of around
$10\,\mathrm{Gyr}$, the large scale picture is dominated by accretion
on to collapsing structures and shocks due to moderate BH
feedback. At this time, a typical Mach number for accretion shocks in
the cosmic web is $\machnum\approx 2$.  The pre-shock gas consists of
streams from voids with temperatures around $4\times10^3\,\mathrm{K}$
(see also Fig.~\ref{fig:tfraction}) and a sound speed of
$c\approx7\,\mathrm{km\,s^{-1}}$, assuming no ionization.  A typical
velocity in the shock frame for the primordial gas streams is
therefore $v=c\machnum \approx 15\,\mathrm{km\,s^{-1}}$. With time the
Mach numbers of shocks around non-linear structures increase. There
appear to be two main reasons for this trend. First, the sound speed
of voids is lowered by the adiabatic expansion of the universe
(although much of this effect is compensated by heating from the UVB) 
and ever larger infall velocities are produced by the
growing masses of objects, and secondly, the BH feedback
processes increase in strength with cosmic time.
 
In particular, between $z=2$ and $z=1$, the BH radio mode
feedback starts driving hot bubbles that resemble Sedov--Taylor blast
waves and feature Mach numbers of several tens in a significant
fraction of the simulation volume.  These shocks become even stronger
at redshift zero when they run into cool and low-density voids, with
peak Mach numbers of several hundred. We can estimate the speed of a
$\machnum=200$ blast wave by neglecting the speed of the pre-shock gas
in the lab frame and assuming a void temperature of
$10^3\,\mathrm{K}$. This temperature corresponds to a sound speed of
$c\approx3.4\,\mathrm{km\,s^{-1}}$ (assuming a mean molecular weight of
neutral gas, for simplicity), and we obtain for the velocity of the
shock $v\approx 700\,\mathrm{km\,s^{-1}}$.  This value is consistent with
the velocities inferred by \citet{HAIDER_2015}, who compared the
diameter of the hot bubbles at different redshifts and reported
expansion speeds of $500-1000\,\mathrm{km\,s^{-1}}$. 

The bottom panels of Fig.~\ref{fig:fullbox} show the energy
dissipation at shocks, which depends on the Mach number,
the pre-shock density, and the pre-shock sound
speed. The latter two quantities are high in the interior of
non-linear structures and the total energy dissipation at shocks is
typically dominated by these locations \citep{RYU_2003}.  It can be
seen that in the Illustris simulation a high energy dissipation rate
is present inside the densest structures between $z=2$ and $z=1$,
however it seems that at $z=0$ the bulk of the thermalization happens
in more extended regions.  Quantifying these qualitative observations
is one of the main goals of this work and will be pursued in subsequent
sections.

Large-scale feedback shocks which are omnipresent at $z=0$ have
the effect of converting BH radio mode feedback energy into
thermal energy of the gas.  In this way, an extended warm hot
intergalactic medium (WHIM) gas phase is created, with temperatures
between $10^5\,\mathrm{K}$ and $10^7\,\mathrm{K}$.  In
Fig.~\ref{fig:tfraction}, we plot the gas volume fraction per
logarithmic temperature bin as a function of temperature for different
redshifts. At all times, void regions occupy the largest volume in the
simulation, and at $z=2$ the distribution peaks around a void
temperature of $5\times10^3\,\mathrm{K}$. With increasing time, this
peak shifts towards lower temperatures due to the expansion of the
universe and the associated adiabatic gas cooling, reaching
$10^3\,\mathrm{K}$ at $z=0$.  Interestingly, the volume distribution
develops a bimodality at late times due to a monotonically increasing
fraction of gas with temperatures above $10^5\,\mathrm{K}$, revealing
the extended WHIM created by the BH radio mode
feedback. \mybf{In non-radiative simulations this bimodality is
not present, as indicated by the blue dashed line.
In Illustris-1, the gas phase with $T>10^5\,\mathrm{K}$ occupies
around $40\%$ of the volume at $z=0$. Furthermore, it
contains more than $60\%$ of the total baryonic mass
\citep{HAIDER_2015}.}

An important effect of AGN feedback from supermassive BHs is
the quenching of star formation in massive systems at late times.  In
fact, the parameters of the adopted radio mode feedback model have
been set in an attempt to match observations of the galaxy stellar
mass function \citep[e.g.][]{BERNARDI_2013} and the cosmic star
formation rate density \citep[e.g.][]{BEHROOZI_2013}. Our results
suggest that it would be potentially also very constraining to
consider the effects of AGN feedback on the predicted WHIM properties.
Unfortunately, this gas phase is difficult to observe due to its low
density and comparatively high temperature, resulting in low emission
and absorption efficiencies, limiting at present the power of this
approach.

Nevertheless, the strong Sedov--Taylor blast wave like shocks due to
BH feedback originate in clusters and groups, where the gas
density is high. In this environment, as we will showcase in
Section~\ref{sec:morphologies}, these shocks can have higher Mach
numbers and larger dissipation rates than cluster merger shocks.  The
latter can be observed in the local Universe in a few cases directly,
and in many others as radio relics. 
Moreover, shocks associated with AGN have been observed
and identified \citep[e.g.][]{NULSEN_2005, NULSEN_2005b, JETHA_2008,
BLANTON_2009, GITTI_2011, CAVAGNOLO_2011, RANDALL_2011}, however,
they are typically of low strength with Mach numbers $\machnum \lesssim 2$ and 
located relatively close to the cluster centre, at distances of at most
$200-300\,\mathrm{kpc}$.
To our knowledge, no strong
shocks with properties similar to the AGN feedback shocks in the
Illustris simulation have been seen in observations thus far.  Their
absence hence indicates that the radio mode feedback channel is too
strong in the simulation model.  
This shortcoming of the simulation
has also been pointed out by other authors based on other lines of
evidence. In particular, as a result of the redistribution of baryons
due to the overly strong AGN feedback, the baryon fraction within
clusters and groups at $z=0$ is underpredicted compared to
observations \citep{GENEL_2014,HAIDER_2015}.

\begin{figure*}
\centering
\includegraphics{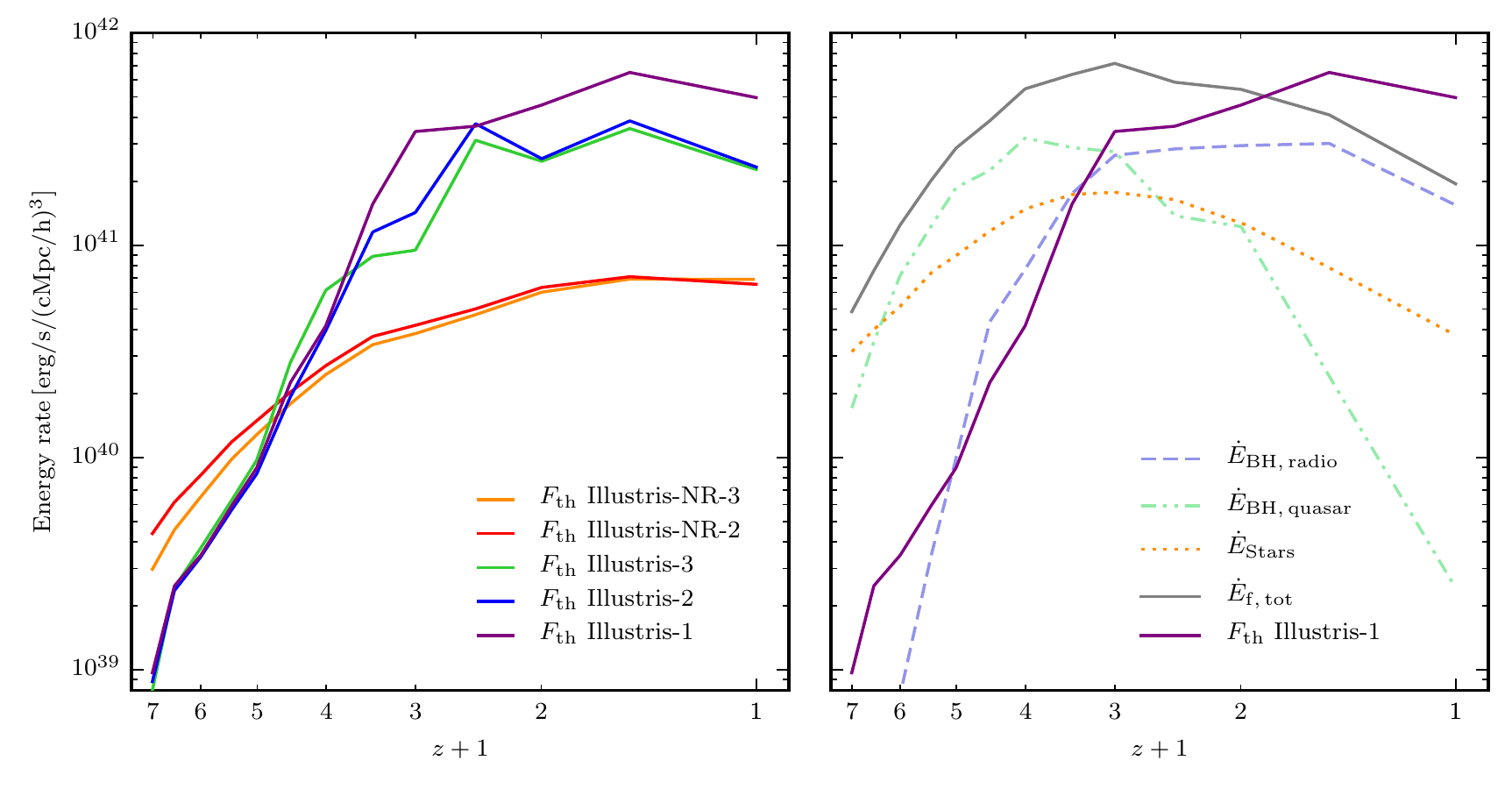}
\caption{{\em Left-hand panel:} total energy dissipation at shocks as a
  function of redshift for the different Illustris runs.  At late
  times, the thermalization in the full physics runs is higher by a
  factor of around 4-8 compared to the non-radiative runs. Moreover,
  there is good agreement for the runs with resolution levels 2 and 3,
  whereas the full physics Illustris-1 run shows a stronger
  dissipation at low redshifts.  This effect presumably originates
  from an increased BH feedback activity in the highest
  resolution run.  {\em Right-hand panel:} energy dissipation at shocks in
  the Illustris-1 simulation compared to the energy output of
  different feedback channels. For redshifts $z>1$, the dissipation
  stays well below the total energy dumped into feedback processes of
  stars and BHs, indicating that not all feedback channels
  produce energetic shocks. On the other hand, there exists a strong
  correlation at redshifts $z<4$ between the energy release by BHs
  in the radio-mode state and the energy rate dissipated at
  shocks.  }
\label{fig:energy_comparison}
\end{figure*}

\subsubsection{Mach number distributions}

In Fig.~\ref{fig:surface_energy}, we show the cumulative shock surface
(left-hand side panel) and the energy dissipation at shocks
(right-hand side panel) as a function of Mach number for different
redshifts. The shock surface is normalized by the 
simulation volume, and it has therefore units of an inverse length-scale.

For $z\ge 2$ by far most of the shocks occur between voids
and filaments of the cosmic web. The shock surface within haloes is
insignificant compared to the total area of shocks, and shocks in
the WHIM only contribute non-negligibly for $z<2$ \mybf{(see also Section
\ref{subsub:envtime}).} Therefore, the shock surface area
is dominated by external shocks.  The red curve ($z=6$)
indicates the distribution of shocks on the eve of reionization, which in
Illustris happens almost instantaneously and is modelled by a uniform
time-dependent ionizing background following \citet{FAUCHER_2009}.

The shock distribution is relatively broad and most of the shocks have
Mach numbers around $\machnum=10$. However, there are also very strong
external shocks present with Mach numbers of up to
$\machnum=1000$. These highly supersonic flows are found between the
coldest voids and strong filaments or halo outskirts.  During the
process of reionization the temperature in voids reaches
$\sim10^4\,\mathrm{K}$ and the distribution is therefore shifted
towards lower Mach numbers.  

\mybf{Naively one expects that many weak shocks between
voids and the cosmic web disappear during reionization, 
since the sound speed in the corresponding gas streams exceeds
their velocities.
However, we do not find that this is the case; 
the Mach number of a typical external shock 
merely decreases from $\machnum\approx 10$ to $\machnum\approx 2$.
We assume that these shock waves can be sustained due to an increase
of their pre-shock velocities, which originates from the 
pressure driven thermal expansion of the filaments during the
reionization heating.}

\mybf{The total shock surface area
increases between $z=6$ and $z=5.5$ (given by the integrals of the curves)
by around 7\%.} Thereafter, it decreases monotonically.  We interpret
this finding as the creation of a small population of newly created
shock waves as the gas responds to the sudden heating during
reionization. Note however that this mechanism may be strongly
alleviated for a more realistic, gradual reionization history that is
stretched out in time, similar to the models arising from radiative
transfer methods applied to the Illustris simulation
\citep{BAUER_2015}. 

After reionization ($z=5.5$ to $z=0$), the shock surface of low Mach number
shocks progressively decreases, whereas the surface area of high Mach number
shocks grows monotonically. This behaviour can be understood if one
recalls that the shock surface statistic is dominated by accretion
shocks from voids on to the cosmic web. As voids evolve they expand
and merge into larger voids (see Fig.~\ref{fig:fullbox}), which tends
to decrease the abundance of low Mach number shocks. At the same
time, their gas becomes colder with time \citep[disregarding
scenarios such as heating by TeV
blazars, see][]{Broderick2012, PFROMMER_2012},
giving rise to a stronger contrast between
voids and the cosmic web, and therefore stronger shocks.

Interestingly, $\machnum \approx 4.2$ is a peculiar Mach number of the
universe; the surface area of shocks with this strength stays roughly
constant after reionization.  At $z=0$, the total shock surface density
reaches a value of
$8.5\times 10^{-1}\,\mathrm{Mpc}^{-1}$ (integral of
the blue curve).  This number is around $3.4$ times higher than what
we have found for Illustris-NR-2 \citep{SCHAAL_2015}.  However, in the
non-radiative run a large fraction of weak shocks on to the cosmic web
is lost by the post-processing reionization modelling in form of a
global temperature floor. We therefore argue that a comparison to
Illustris-NR-2 without a temperature floor is more meaningful. In this
case we find that in Illustris-1 the total shock surface area
is about $1.4$ times higher.

The energy dissipation at shocks as a function of Mach number is shown
in the right-hand side panel of Fig.~\ref{fig:surface_energy}. This
shock property is strongly dependent on the environment and high for
regions with high densities ($f_\text{th}\propto\rho_1$).  Although
the shock Mach numbers decrease during the process of reionization,
the energy dissipation increases significantly.  This can be partly
attributed to the triggering of new shocks, but also to the
modification of existing ones. We expect that non-linear structures,
such as filaments, dynamically respond to the heating by slightly
expanding.  In this way, the inflowing kinetic energy can be
increased.

Throughout cosmic evolution, the shock distribution remains broad and
a whole range of Mach numbers contribute comparably. For example, 50\%
of the thermalization happens at shocks with strengths below
$M\le6-10$, the other half in higher Mach number shocks.  As opposed
to the low Mach number shocks in the surface statistic, the low Mach
number shocks in the energy statistic are shocks well inside
non-linear structures, which we refer to as internal shocks.  This
distribution differs substantially from what is found in non-radiative
simulations, where 50\% of the dissipation occurs in $\machnum<4$
shocks, and high Mach number shocks are present but do not contribute
significantly to the total dissipation. 

Non-radiative simulations show
a bimodality in the energy statistics, consisting of low Mach number
internal shocks which dissipate most of the energy, and high Mach
number external shocks processing cold primordial gas \citep{RYU_2003,
  VAZZA_2011}.  Since it can be a source of confusion we want to point
out that the bimodality in the energy statistics of non-radiative
simulations can not be directly mapped to the surface distribution
of shocks.  While high Mach number shocks are external shocks, most of
the external shocks actually have low Mach numbers \citep[see also
Fig.~12 in][]{VAZZA_2009}.  Moreover, the adopted modelling of
reionization can erase weak shocks on to the cosmic web, hence a
self-consistent interpretation can ultimately only be achieved if
reionization is followed self-consistently during the simulation, as
is the case for Illustris-1, albeit with a simplified reionization
history.

In the dissipation statistics we measure for the full physics
Illustris-1 run, the appearance of energetic high Mach number shocks
prevents the development of a bimodality. This difference to the
non-radiative simulations is striking, and as we will show later on,
the additional shocks can be attributed to feedback processes.  

\subsubsection{Total energy dissipation at shocks}

In Fig.~\ref{fig:energy_comparison}, we compare the total energy
dissipation at shocks across cosmic time for the full physics and
non-radiative runs at different numerical resolution.  For these
simulations the shock finder gives consistent results with respect to
resolution.  In general, the global dissipation rate shows a shallow
but strictly monotonic increase at early times before a saturation
sets in.  In the full physics runs, a strong increase in the total
dissipation rate at shocks can be seen during reionization
$z\approx 6$, which, as discussed above, presumably originates from
the triggering of new shocks and the dynamical response of non-linear
structures to the heating.  The next four data points between $z=5.5$
and $z=4$ indicate a power law behaviour, before the energy rate
steepens and slightly diverges for the runs with different
resolutions. We suggest that the latter feature marks the time when
feedback processes from stars and BHs become important for the
total energy dissipation at shocks. 

From $z=4$ to $z=2$, the energy rate rises by more than an order of
magnitude and bends sharply thereafter. This sharp bend coincides with
the creation of the extended WHIM at $z=2$ (recall
Fig.~\ref{fig:tfraction}), indicating that internal shocks are erased
in this process. At $z=0$, the simulations Illustris-1, Illustris-2,
and Illustris-NR-2 reach final values of
$1.7\times 10^{41}\,\mathrm{erg}\,\mathrm{s}^{-1}\,\mathrm{Mpc}^{-3}$,
$8.1\times 10^{40}\,\mathrm{erg}\,\mathrm{s}^{-1}\,\mathrm{Mpc}^{-3}$,
and
$2.3\times 10^{40}\,\mathrm{erg}\,\mathrm{s}^{-1}\,\mathrm{Mpc}^{-3}$,
respectively.  In either case, due to the creation of feedback shocks
the dissipation in the full physics runs at late times is higher by a
factor of several compared to the non-radiative runs.

While the Illustris-3 and
Illustris-2 runs give fairly similar results, the dissipation at
shocks in the highest resolution run is considerably larger at low redshifts. As we
show in Appendix~\ref{sec:resolution_study}, the difference originates
from high Mach number shocks and correlates with a larger amount of
feedback energy released in Illustris-1. Most importantly, between
$z=2$ and $z=0.5$ the BH radio mode feedback is stronger by a
factor of around 2. This indicates that BH accretion histories in Illustris
are resolution-dependent, and more gas can be funnelled to central
regions if the gas dynamics are resolved with a higher resolution.

In order to put the values we measured for the energy dissipation at
shocks in Illustris-1 into context, we compare the dissipation rates
to other characteristic energy rates in the simulation.
Specifically, we want to compare the kinetic energy dissipated at
shocks with the energy output of BHs and stars.  The energy
release by AGN in Illustris-1 scales with the accretion rate
$\dot{M}_\text{BH}$ on to the BH and is given by
$\dot{E}_\text{BH}=\epsilon\epsilon_\mathrm{r}\dot{M}_\text{BH}c^2$.
Here, $\epsilon_r=0.2$ is the radiative efficiency and
$\epsilon\in\{\epsilon_m, \epsilon_f\}$ the efficiency with which the
feedback couples to the surrounding gas. Depending on the accretion
rate the BH is either in the radio feedback mode
($\epsilon=\epsilon_m=0.35$) or quasar feedback mode
($\epsilon=\epsilon_f=0.05$).  The corresponding criteria are
$\dot{M}_\text{BH}<\chi_\text{radio}\dot{M}_\text{Edd}$ and
$\dot{M}_\text{BH}\ge\chi_\text{radio}\dot{M}_\text{Edd}$,
respectively, where $\chi_\text{radio}=0.05$ denotes the radio
threshold and $\dot{M}_\text{Edd}$ the Eddington accretion rate.
Moreover, the injected SNe feedback energy can be inferred by
multiplying the cosmic star formation rate density with the SNII
energy per stellar mass, which in Illustris-1 has the value of
$\text{egy}_\mathrm{w}=1.09\times 1.73 \times 10^{-2} \times
10^{51}\,\mathrm{erg}/\msol$ \citep{VOGELSBERGER_2014c}.

In the right-hand panel of Fig.~\ref{fig:energy_comparison}, we
present the kinetic energy dissipated at shocks in Illustris-1 as well
as the energy output of the different feedback channels.  The grey
lines indicate from top to bottom the BH radio mode energy
rate, the BH quasar mode energy rate, the feedback energy rate
dumped into stellar winds, and the sum of the feedback energy
rates. Several interesting aspects of the adopted galaxy formation
model become apparent.  First of all, due to the co-evolution of BHs 
and galaxies, the BH feedback increases at early times in
parallel with an increasing SNe energy rate.  Later on, the
BH accretion rate declines, resulting in a decrease of the
BH quasar mode activity and an increase in radio mode
activity. The latter trend coincides with an onsetting decline in the
star formation rate density, highlighting the importance of the radio
mode feedback for quenching star formation in the full physics runs.

For $z>0.5$, the shock energy dissipation stays well below the total
feedback energy, indicating that not all feedback channels drive
energetic shocks. Especially the AGN quasar-mode feedback and
SN feedback offer more net energy than what is detected by
shocks for $z>3$. The former heats central regions of galaxies, and
presumably much of its energy is radiated away in cooling processes
before it can drive energetic shock waves.  Galactic winds on the
other hand are often launched inside hot haloes where the sound speed
is high, making them unlikely to appear as supersonic flows, and in
addition, they also suffer from strong radiative cooling losses.
Nevertheless, as we will show in Section~\ref{sec:morphologies}, there
are some galaxies which show quasar mode feedback shocks and launch
supersonic winds.

For redshifts $z<4$, there is a remarkable correlation between the
radio mode AGN feedback rate and the dissipation measured at shocks.
We interpret this finding as an indication that this mode is not only
able to drive high Mach number and energetic shocks, as seen in
Fig.~\ref{fig:fullbox}, but that these strong feedback blast waves
also contribute significantly to the integrated energy statistics at
late times.

Hence, it is the BH feedback in the full physics runs which
can explain a significant fraction of the overall higher energy
dissipation compared to the non-radiative runs.  Another source of
additional energy dissipation at shocks in the full physics runs could
be the presence of gas cooling via radiative processes.  In this case,
dense streams are formed as well as cold, concentrated gas blobs,
which are difficult to disrupt and create ram-pressure shocks absent
in the non-radiative simulations.

\subsection{Environmental dependence of the shock statistics}
\label{sec:environmental_statistics}

\subsubsection{Environmental dependence at redshift zero}
\label{subsub:envzero}

\begin{figure*}
\centering
\includegraphics{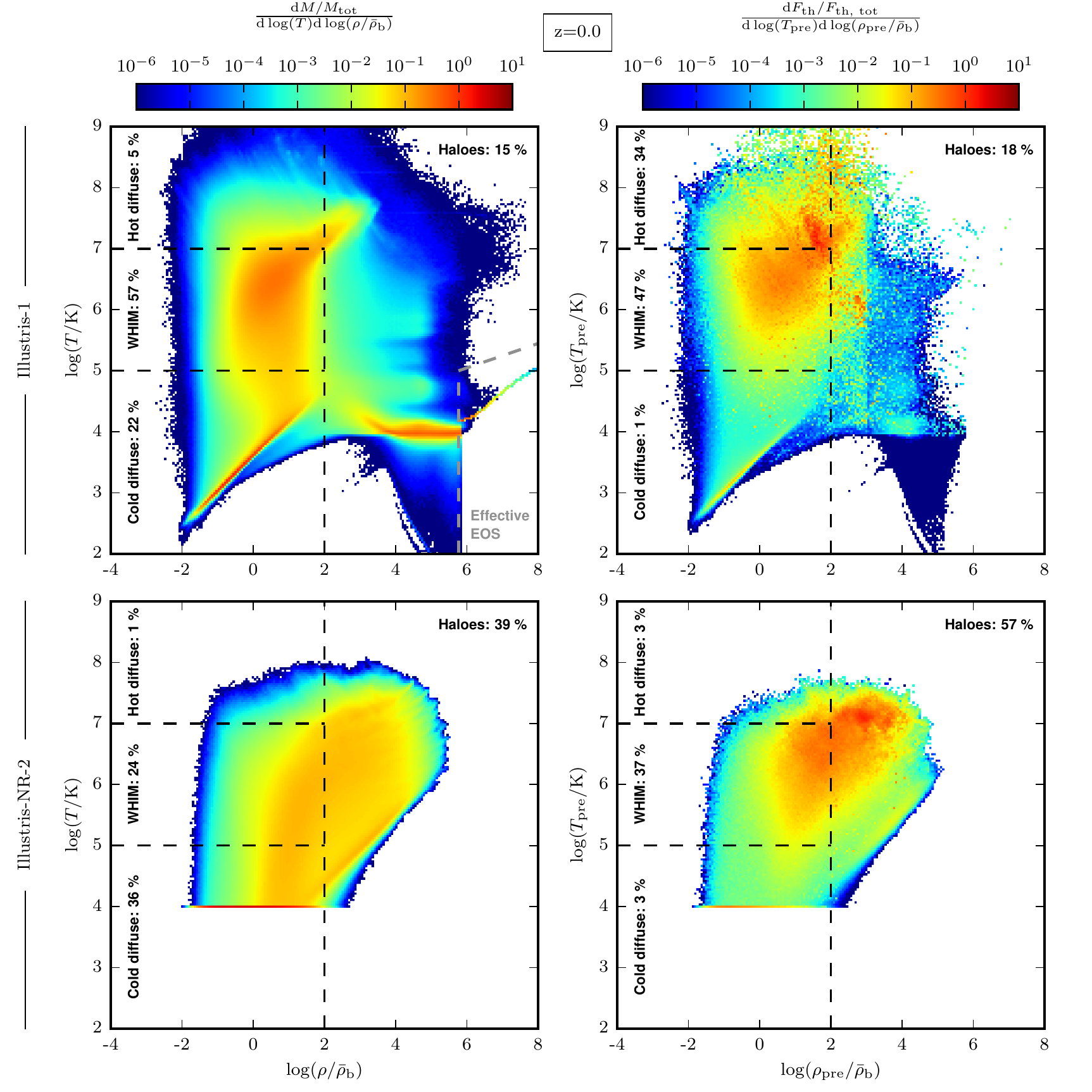}
\caption{Gas density-temperature phase-space diagrams at redshift
  $z=0$ weighted with gas mass (left-hand side panels) and energy
  dissipation at shocks (right-hand side panels). The latter are
  constructed with respect to pre-shock gas quantities.  The top
  panels show results for the full physics Illustris-1 simulation, and
  the bottom panels for the non-radiative Illustris-NR-2 run.  In the
  full physics run, BH feedback processes transfer gas mass
  from haloes to the ambient medium, creating an extended WHIM region
  which contains around $60\%$ of the gas mass at $z=0$. Similarly, a
  large fraction of the energy dissipation is contributed by shocks in
  the WHIM, which dominate the dynamics of this gas phase. Remarkably,
  a considerable fraction of dissipation is also present in the hot
  diffuse phase, where relatively little mass resides.  }
\label{fig:histograms}
\end{figure*}

\begin{figure*}
\centering
\includegraphics{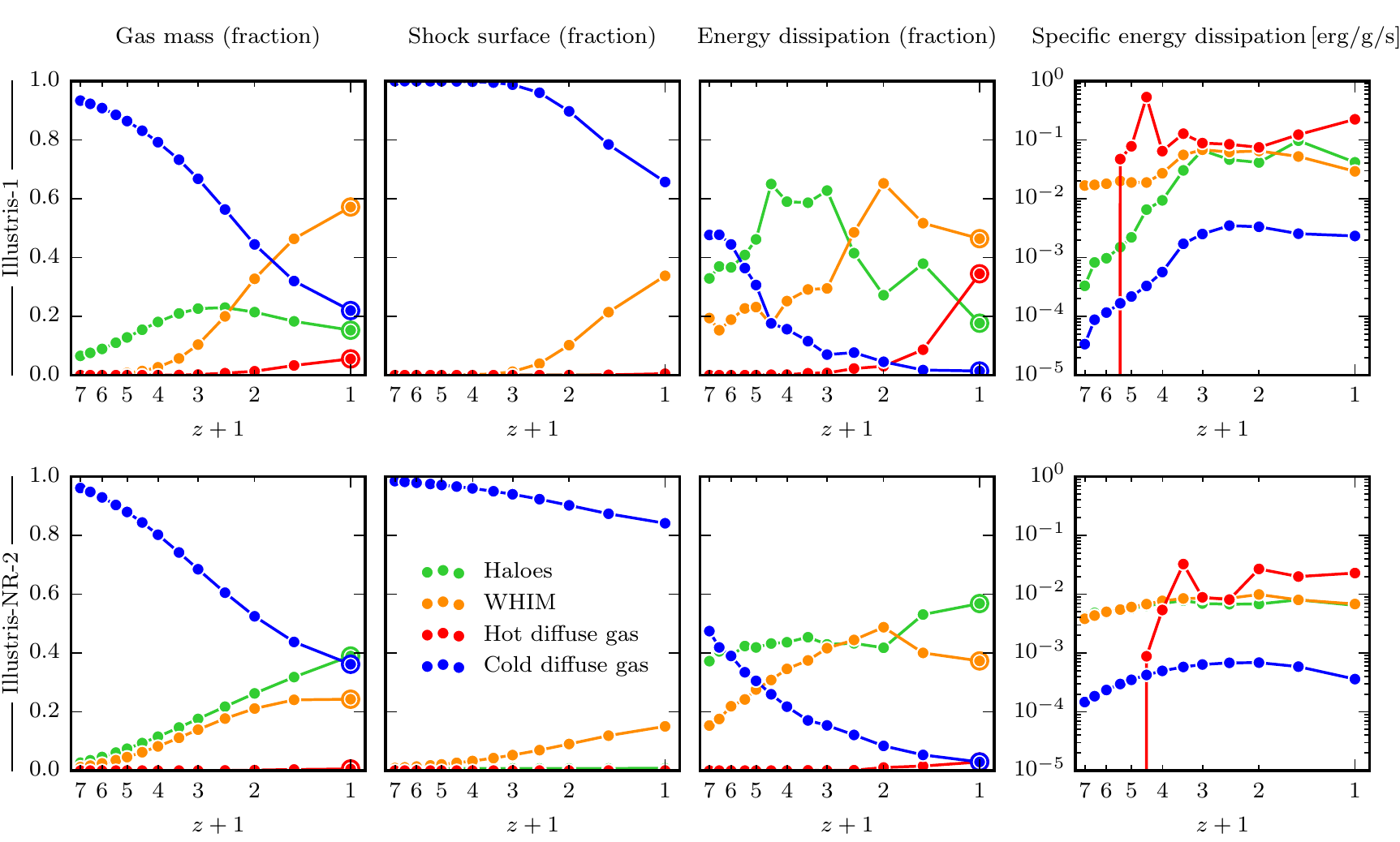}
\caption{Time evolution of the contribution of different environments
  to, from left to right, the gas mass fraction, the shock surface fraction,
  the dissipation at shocks, and the dissipation per unit mass.  In the
  full physics as well as in the non-radiative run most of the shock surface area
  resides in the cold diffuse phase, especially at high redshifts.
  Between $z=3.5$ and $z=2$, shocks inside haloes play the most
  important role in thermalising energy in Illustris-1, and their
  relative contribution is significantly higher compared to the
  non-radiative run. This clearly demonstrates the impact of
  non-radiative physics, which is most important in this
  environment. For $z<1.5$, most of the energy is dissipated in the
  extended WHIM phase and even the hot diffuse phase contributes
  significantly.  The fact that for the latter the shock surface area
  is very small indicates that this environment is
  created and heated by individual highly energetic shocks.
  Interestingly, except for the cold diffuse phase, the different
  environments have a similar and roughly constant specific energy
  dissipation for $z<3$.}
\label{fig:time_evolution}
\end{figure*}

In the previous section we have analysed the occurrence of shocks in the
Illustris universe and the associated energy dissipation in a global
sense as a function of Mach number and redshift. We are now turning to
an investigation of the importance of different environments in the
thermalization processes during cosmic evolution. For this purpose, we
define four disjoint regions based on the gas temperature $T$ and
overdensity
$\delta_\mathrm{b}=(\rho_\mathrm{b}-\bar{\rho}_\mathrm{b})/\bar{\rho}_\mathrm{b}$,
where $\bar{\rho}_\mathrm{b}$ is the mean baryon density of the
universe.  Those environments are cold diffuse gas
($T<10^5\,\mathrm{K}$ and $\delta_\mathrm{b}<100$), hot diffuse gas
($T\ge10^7\,\mathrm{K}$ and $\delta_\mathrm{b}<100$), the WHIM
($10^5\le T/\mathrm{K}<10^7$ and
$\delta_\mathrm{b}<100$)\footnote{\label{foot:whim} Some authors
  define the WHIM environment as $10^5\le T/\mathrm{K}<10^7$,
  independent of the gas overdensity. However, in the case of the full
  physics run our definition is very similar since by far most of the
  gas in this temperature range is located outside of haloes and
  therefore has overdensities below $100$, as can be seen in
  Fig.~\ref{fig:histograms}.}, and haloes ($\delta_b\ge100$).
  
Fig.~\ref{fig:histograms} presents the distributions of gas mass
(left-handed panels) and energy dissipation at shocks (right-handed
panels) within the different regions at $z=0$.  The histograms that
are weighted with energy dissipation are constructed with respect to
pre-shock quantities, and all histograms are normalized to unity.  The
panels on the top and bottom show results obtained for the Illustris-1
and Illustris-NR-2 simulations, respectively.  Note that for the
non-radiative run, reionization is modelled in post-processing by
simply adopting a temperature floor of $10^4\,\mathrm{K}$. A large
fraction of the mass and dissipated energy of the cold diffuse phase
resides in the corresponding individual temperature bin.

Several features in the mass-weighted density-temperature phase
diagram of Illustris-1 are apparent (top panel on the left-hand side).
In the cold diffuse environment, most of the gas follows a mean
equation of state, which is a power law governed by the net effect of
photoionization-heating and adiabatic cooling due to the Hubble
expansion \citep{HUI_1997}.  Gas inside the WHIM is characterized by
an increased temperature due to shock heating. A correlation can also
be seen inside this phase \citep{DAVE_2001}, which is however less
tight and exhibits more scatter. The latter observation is expected
due to the variety in number, origin, and properties of shocks
governing the dynamics of the WHIM.

Most of the gas above $\delta_\mathrm{b}=100$ is in a cool phase
around $10^4\,\mathrm{K}$ within galaxies, where photoionization
heating and radiative cooling are close to equilibrium
\citep{VOGELSBERGER_2012}.  Condensed gas above the star formation
threshold (vertical grey dashed line) follows an effective equation of
state, which represents a subgrid model for a pressurized ISM
consisting of a hot and cold phase \citep{SPRINGEL_2003}.  The bulk of
gas mass in the Illustris simulation at redshift zero is contained in
the WHIM ($\approx 60\%$), followed by the cold diffuse phase
($\approx 20\%$), and gas within haloes ($\approx 15\%$).

This gas mass distribution differs significantly from the distribution
of the non-radiative Illustris-NR-2 run, in which $\approx 40\%$ of
the gas is contained within haloes, and only $\approx 25\%$ inside the
WHIM at $z=0$. The difference arises mainly due to BH feedback
processes in Illustris-1, which transfer gas from haloes to the
ambient medium, creating an extended WHIM phase. However, compared to
observations the gas mass fractions inside massive haloes
($M\gtrsim10^{13}\,\msol$) in Illustris-1 at the present epoch appears
too low by a factor of several \citep{GENEL_2014}, indicating that the
redistribution is too efficient.  For a comprehensive analysis of the
mass distribution in the Illustris simulation and comparison to
previous work as well as observations, we refer the reader to
\citet{HAIDER_2015}.

In the non-radiative simulations, for which radiative cooling and
feedback is absent, the temperature evolution of the gas is dominated
by shock heating on the one hand, and adiabatic cooling due to the
Hubble flow on the other hand. Most of the energy dissipation at
shocks at $z=0$ takes place internally within haloes and filamentary
structures, summing up to around $60\%$ of the total dissipation.
Those internal shocks typically have high pre-shock sound speeds and
therefore low Mach numbers ($2\lesssim\machnum\lesssim 4$).  However,
they dissipate strongly since the inflowing kinetic energy flux is
large due to the involved high densities.  The second relevant
environment is the WHIM, where merger and accretion shocks contribute
almost $40\%$ to the total thermalization by shocks.  A more extended
discussion of shock statistics and morphologies for the Illustris-NR-2
simulation can be found in \citet{SCHAAL_2015}.

We note that for the full physics runs we exclude cells from our shock
analysis which are in a region around the effective equation of state,
as indicated by the grey dashed lines in the mass-weighted histogram.
Moreover, we filter inconsistent jumps above $\delta_\mathrm{b}=1000$,
and this density threshold can be seen slightly as an edge in the energy
dissipation weighted histogram. 
Its visibility depends on the choice for the parameter $f$ of our shock finding
algorithm, however, as will be shown in Section~\ref{sec:methodology_variations},
the choice does not influence the statistics significantly.

In the Illustris-1 simulation, about $80\%$ of the dissipated shock energy
is found inside the WHIM and the hot diffuse phase.  Compared to the
non-radiative run, differences in the WHIM potentially arise
from photoionization heating, radiative cooling, and feedback
processes from stars and BHs. However, heating due to
photoionization increases the gas temperature only to around
$10^4\,\mathrm{K}$, which is well below the temperature of the WHIM.
Moreover, radiative cooling plays only a subdominant role in the
evolution of the WHIM \citep{DAVE_2001}, since the bulk of the gas has
low density. 

SN feedback can potentially dump thermal energy
into these gas phases, depending on the implementation of this
feedback channel and the numerical setup. In Illustris-1, winds are
launched in star-forming regions within galaxies and couple to the gas
in the near vicinity. These winds can drive shocks and increase the
thermal energy of gas within the halo, as shown in
Section~\ref{sec:morphologies}. However, they typically do not reach
the virial radius. Consequently, the influence of SN feedback
on the WHIM in Illustris-1 can be neglected.  We therefore argue
that the increased importance of the WHIM and the hot diffuse phase in
thermalising kinetic energy originates from BH feedback.  

Compared to these phases, the cold diffuse phase is not important in
this respect due to its low density. The dissipation within haloes
amounts to $15\%$, which is low compared to the non-radiative run.
Note however that in Illustris-1 the overall energy dissipation at
late times is higher by a factor of around 8, so compared to the
non-radiative simulation the total energy dissipation within 
haloes is higher.

\subsubsection{Environmental dependence across cosmic time}
\label{subsub:envtime}

In Figure~\ref{fig:time_evolution}, we compare several properties of
the different environments across cosmic time.  From left to right, we
show the gas mass fraction, the shock
surface fraction, the energy dissipation rate, as well as the specific
dissipation rate.  For the shock related quantities the environments
are defined with respect to the pre-shock region, and the circled data
points at $z=0$ correspond to the integrated phase diagrams of
Fig.~\ref{fig:histograms}.  Results for the full physics run are shown
in the upper panels, while the lower panels indicate results obtained
for the non-radiative run.

As expected, the gas mass fractions in the non-radiative run show a
monotonic evolution, in which mass is transferred from low to high
density regions.  In contrast, BH feedback redistributes
baryons in Illustris-1 from haloes to the WHIM and the hot diffuse
phase, decreasing the gas mass fraction of haloes at late times to a
value below $20\%$.  Moreover, in the full physics run a steeper
decrease of the mass in the cold phase is present for $z<2$.  This can
be explained by the fact that the WHIM does not only grow in mass, but
also becomes more spatially extended, as we have seen in
Fig.~\ref{fig:tfraction}.

The shock surface fraction clearly demonstrates that in both kinds of
simulations the shock heating of gas in the cold diffuse phase
produces the largest contribution to the total shock surface area. 
Most of these shocks are external
shocks, created when pristine gas from voids accretes on to non-linear
structures. Only for $z<2$ does the WHIM contribute a significant fraction of 
shock surface area, and furthermore, the cumulative surface area of 
shocks in haloes and in the hot phase are negligible in comparison.

While measuring the shock surface only reveals the locations of shocks, 
the energy dissipation quantifies the impact on the
thermal history of the gas.  Comparing Illustris-1 to Illustris-NR-2
in this respect reveals that between redshifts $z=3.5$ and $z=2$ a
higher fraction of kinetic energy is thermalized within haloes in the
full physics run.  As can be seen from the left-hand panels, this
correlates with a higher halo gas mass fraction in Illustris-1 at
early times due to radiative cooling.  With more available mass and
denser streams more energy can potentially be dissipated.  On the
other hand, the increase within haloes is abrupt and also coincides
with the onset of the BH radio mode feedback at
$z\approx 4-3$. Due to the subsequent mass transfer and
expansion of the WHIM, a higher fraction of energy gets dissipated in
this environment. Interestingly, shocks in the hot diffuse phase also
contribute significantly, although there is only a small increase in
the mass fraction and a negligible increase in the shock surface area in
this environment. This suggests heating through few but very energetic
shocks, pointing once more towards the radio-mode blast waves launched
by BHs.

In the panels on the right-hand side we show the specific energy
dissipation, which is the dissipation rate per unit mass. Remarkably,
apart from the cold diffuse phase, all environments have within a
factor of several similar and roughly constant rates for $z<3$.  This
finding is non-trivial, since although the energy dissipation is
proportional to the pre-shock density of the inflowing gas, there is
also a strong dependence on the sound speed and the Mach number.  The
specific dissipation rate in the non-radiative run is around
$10^{-2}\,\mathrm{erg}\,\mathrm{g}^{-1}\mathrm{s}^{-1}$, and for
Illustris-1, we measure a 10 times higher value of approximately
$10^{-1}\,\mathrm{erg}\,\mathrm{g}^{-1}\,\mathrm{s}^{-1}$.  In order
to put these numbers in context we compare them to the power released
in SNe explosions with respect to the mean baryon density of
the universe. This rate can be calculated for Illustris-1 as the
product of the star formation rate density and the SNII energy per
stellar mass,
$P=\text{SFRD}\times\text{egy}_\mathrm{w}\approx0.1\,\msol/\mathrm{yr}/\mathrm{cMpc}^3\times
1.09\times 1.73 \times 10^{-2} \times 10^{51}\,\mathrm{erg}/\msol$,
so that we obtain for the specific power a value of
$P_\mathrm{s}=P/\bar{\rho}_\mathrm{b}=3.2\times
10^{38}\,\mathrm{erg}\,\mathrm{yr}^{-1}\,\msol^{-1}=5.0\times
10^{-3}\,\mathrm{erg}\,\mathrm{g}^{-1}\,\mathrm{s}^{-1}$.
Hence, in Illustris-1, the energy rate per unit mass of shocks is
higher by a factor of 20 compared to the specific energy rate of
stellar feedback.

\section{Shock morphologies across cosmic time}
\label{sec:morphologies}

The aim of this section is to discuss the different shock morphologies
found in the Illustris-1 simulation. We mostly present a qualitative
overview, with the intention to stimulate a more comprehensive
analysis of some of the rich phenomenology exemplified here in future
work.

\begin{figure*}
\centering
\includegraphics{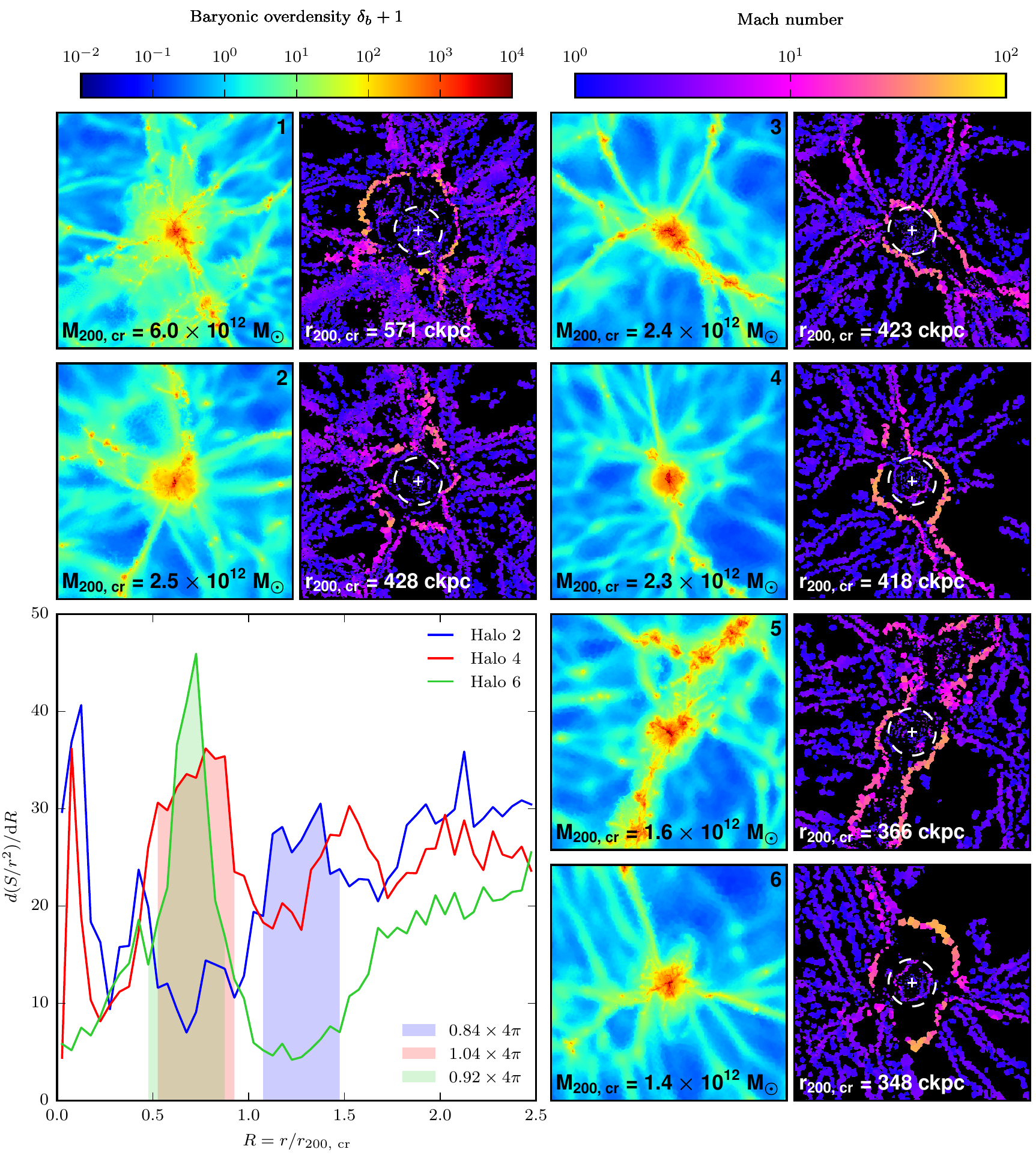}
\caption{Shock environments around some of the largest haloes in the
  Illustris-1 simulation at $z=4$.  Shown are thin projections of the
  baryonic overdensity and the Mach number, with projection depths of
  $0.5 \times \rvir$ and $0.1 \times \rvir$, respectively. 
  \mybf{The white circles correspond to the virial radii, and their
  values are given in comoving kiloparsec at the bottom of the Mach number panels.} 
  In all haloes strong accretion shocks can be seen, which are typically
  located outside of the virial radius.  Moreover, these outer
  accretion shocks can get penetrated by filaments, depending on their
  size and hydrodynamic properties. Interestingly, in several cases a
  second inner accretion shock is formed, especially when the flow
  inside the penetrating filament is smooth (haloes 2, 4, and 6).  The
  lower-left panel shows the shock surface distribution as a function
  of radius for the corresponding haloes. In all three cases, the
  inner accretion shock covers a solid angle of approximately $4\pi$,
  indicating that they are spherical.}
\label{fig:halosz4}
\end{figure*}

\subsection{High-redshift accretion shocks}
\label{sec:highzaccr}

We begin with an investigation of the morphology of accretion
shocks. Figure~\ref{fig:halosz4} shows the shock environment around
some of the biggest haloes in the simulation at $z=4$. As we have seen
in Section~\ref{sec:global_statistics}, the violent BH radio
mode feedback, which operates during low accretion rates, is still
largely absent at this time. Moreover, we do not expect many shocks
from the quasar mode feedback, and the stellar feedback operates on
smaller scales compared to the one under consideration.  We have
checked the environment of the same haloes in the non-radiative
Illustris-NR-2 run (not shown), and find shocks with very similar Mach
numbers and locations. Consequently, the shocks seen in
Fig.~\ref{fig:halosz4} are indeed accretion shocks.

\begin{figure*}
\centering
\includegraphics{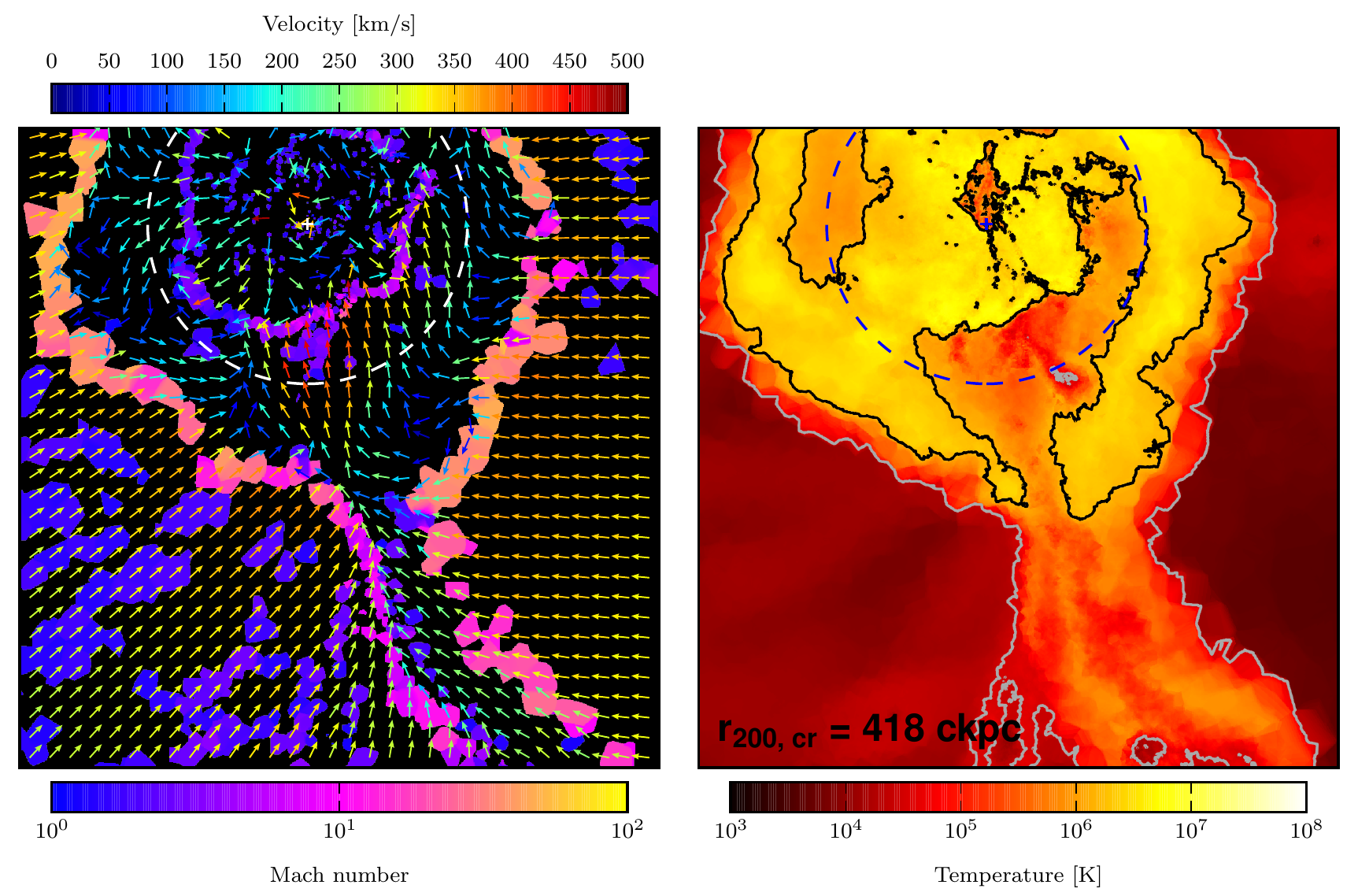}
\caption{Zoom on to a halo with a double accretion shock (halo 4 from
  Fig.~\ref{fig:halosz4}).  The left-hand panel shows a thin projection of
  the Mach number field, as well as the direction and magnitude of the peculiar
  gas velocity field (coloured arrows).  On the right-hand side panel,
  the mass-weighted mean temperature is projected, together with 2D contours at
  $T=3\times10^4\,\mathrm{K}$ (grey) and $T=1.5\times10^6\,\mathrm{K}$
  (black) for visual guidance.  The dashed circles correspond to the
  virial radius, $\rvir=418\,\mathrm{ckpc}$.  A vertical filament
  penetrates the outer accretion shock and channels cold gas into the
  interior.  The cold stream coming from the bottom is accelerated and
  heated at a smaller radius by a strong shock ($\machnum\approx10$).
  Moreover, since the cold phase mixes with gas shock heated at the
  outer accretion shock, a roughly spherical inner accretion shock is
  formed.  }
\label{fig:velocity_zoom}
\end{figure*}

The importance and character of shocks in the formation of gaseous
haloes as well as their role in galaxy formation have been a lasting
topic in the literature.  In the classical two-stage galaxy formation
theory \citep{WHITE_1978} infalling gas gets shock-heated to the
virial temperature before it can cool and settle into a disc, forming
stars inside out.  The pressurized atmosphere of the gaseous halo can
either be created due to thermalization in a strong accretion shock,
or through a succession of smaller shocks inside the halo.  The
scenario in which the gas gets shock heated to the virial temperature
before it cools and fragments into stars is termed the `hot/slow' mode of gas
accretion.  If the cooling time is much shorter than the dynamical
time, which is the case for low-mass galaxies, gas can also be
accreted in a `cold/rapid' mode without being shocked \citep{FARDAL_2001,
  BIRNBOIM_2003}, something that is also present in the original
theory of hierarchical galaxy formation
\citep{White1991}. Furthermore, there may exist a third mode, in which
an outer accretion shock gets penetrated by cold filaments, feeding
the central galaxy with cold and unshocked gas
\citep[e.g.][]{DEKEL_2006, VOORT_2012}.

The haloes shown in Fig.~\ref{fig:halosz4} have masses well above
$\mvir>10^{12}\,\msol$ and are among the most massive haloes at
$z=4$. The specific haloes have been selected for illustrating the
wide variety of occurring shock features. For haloes in this mass 
range, theoretical models predict the existence of stable accretion shocks 
\citep{DEKEL_2006}. The distance of the outer
accretion shock from the halo centre as inferred from non-radiative
simulations is on average around $1.3\times\rvir$ \citep{SCHAAL_2015}.
However, the accretion shock can sometimes also be found significantly
further outside ($\lesssim2\times\rvir$), especially if the halo is in
a non-equilibrium state after a major merger event
\citep{NELSON_2015}.

For all haloes in Fig.~\ref{fig:halosz4}, the presence of an outer
accretion shock is evident, with varying Mach number depending on the
environment. The accretion shock is stronger if gas is infalling
directly from voids, as it is the case for halo 4 in the bottom-right,
and halo 6 in the top-right direction, with Mach numbers up to
$\machnum=100$. The corresponding accretion flows are highly
supersonic due to their low sound speed.  On the other hand, if the
halo is surrounded by weak filaments whose gas is shock heated at the
outer accretion shock, the typical Mach numbers are significantly
lower (e.g. halo 2, $\machnum\approx 10$).  In several locations, we
see that strong filaments are able to penetrate the outer accretion
shock. This preferentially happens for filaments with high gas
densities and velocities, resulting in a large ram pressure
$p_\mathrm{r}\propto \rho v^2$.

Interestingly, although the masses of the haloes are similar, we find
a considerable variety among their interior shock morphologies.
Halo~1 dissipates thermal energy internally by a network of complex
weak shocks.  This network is likely induced by recent merger events,
since it also coincides with an increased outer accretion shock
pointing towards a non-equilibrium system.  Strikingly, in haloes 2, 4,
and 6, a second accretion shock located inside of the outer accretion
shock can be seen.  These shocks can potentially be formed if there is
cold gas from filaments crossing the outer accretion shock unshocked.

The lower left panel of Fig.~\ref{fig:halosz4} shows the radial shock
surface distribution of these haloes, and for each halo the inner accretion
shock can clearly be seen as a peak in the distribution. Moreover, by integrating
the peaks one obtains the solid angles covered by shocks
at the corresponding radii.
We integrate each curve at the location of the inner accretion shock
over an interval of size $0.45\,R$, which accounts
for the deviation from perfect sphericity of the shocks.
The integrals of the inner accretion shocks of haloes 2, 4, and 6 amount to $0.84\times 4\pi$, 
$1.04\times 4\pi$, and $0.92\times 4\pi$, respectively,
indicating that they are largely spherical. 

Haloes 3 and 5 are also penetrated by strong
filaments, but in these cases no inner spherical accretion shocks can
be found. On the other hand, these haloes experience a more clumpy
accretion flow compared to halo 4 and 6, which might prevent the
formation of the inner spherical shock.

The existence of inner and outer accretion shocks is consistent with
recent work by \cite{NELSON_2015}, who analysed gas accretion on to
$10^{12}\,\msol$ haloes at $z=2$ by means of cosmological hydrodynamic
zoom simulations\footnote{Although this study also uses the {\small
    AREPO} code, very different simulation physics have been used
  compared to the Illustris runs.}.  In addition to a strong virial
shock at around $1.3\times\rvir$, they report for their analysed
haloes that inflows which are not shocked at the virialization
boundary are significantly heated inside the viral radius.  This
happens at a distinct distance from the centre, typically at
$\lesssim0.5\times\rvir$, and only thereafter the gas cools and
accretes on to the central galaxy in the examined systems.

Fig.~\ref{fig:velocity_zoom} shows a zoom on to halo 4 that focuses on
projections of the Mach number and temperature field.  Additionally,
in the left-hand panel the direction and magnitude of the peculiar gas velocity
is visualized by the orientations and colours of arrows, respectively.  The
halo has a virial mass of $\mvir=2.3\times 10^{12}\,\msol$ and is
surrounded by a strong outer accretion shock with Mach numbers of
$\machnum=40-50$. Such shock strengths are expected for relatively
cool accretion flows from voids and weak filaments.  Moreover, gas
inside the highlighted filament is dense and fast enough to penetrate
this outer shock and thermalize further in, where the halo pressure is
higher. The accretion flow from the filament gets accelerated towards
the potential minimum and reaches a velocity of around
$500\,\mathrm{km}\,\mathrm{s}^{-1}$ relative to the halo center.  With
a pre-shock temperature of $10^5\,\mathrm{K}$ and assuming full
ionization, this configuration corresponds roughly to a Mach number
$10$ shock, consistent with the value obtained by the shock finder.
Moreover, given the location inside the virial radius, this shock can
be considered very strong.

\begin{figure*}
\centering
\includegraphics{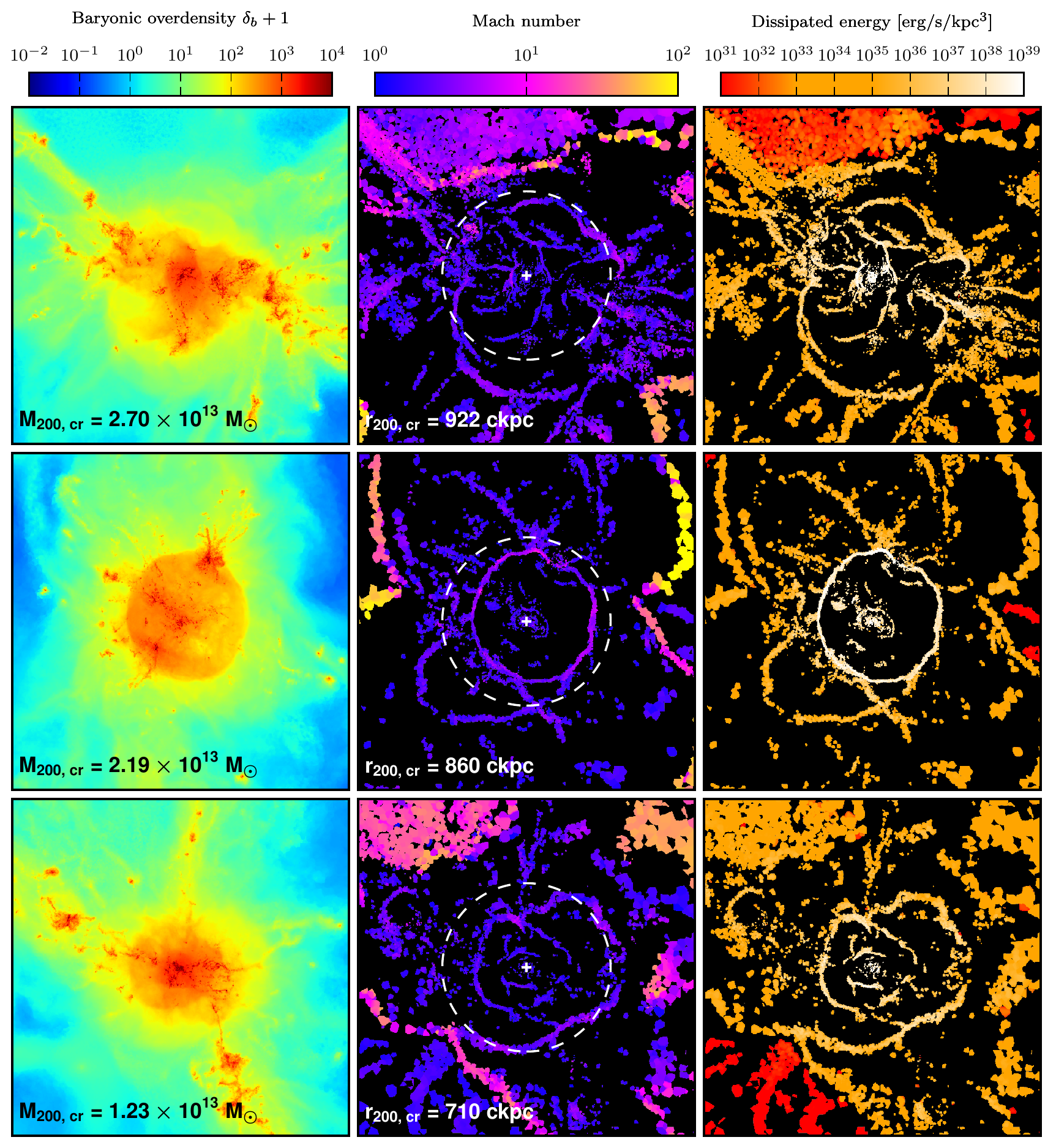}
\caption{Zoom on to some of the most massive galaxy clusters at $z=2$ in
  Illustris-1. Shown are projections of the baryonic overdensity, the
  Mach number, and the energy dissipation, with depths of
  $0.5 \times \rvir$ for the density, and $0.1 \times \rvir$ for the
  shock related quantities. In the halo of the first row, merger activities can be
  seen which induce a complex network of internal shocks. These shocks
  as well as the stimulated supersonic turbulence inside the cluster are
  potential cosmic ray particle accelerators. The BH of the
  halo displayed in the middle row recently released a large 
  amount of thermal energy,
  resulting in an energetic Sedov--Taylor like blast wave feedback
  shock. The halo at the bottom presumably underwent a major merger recently,
  as revealed by the arc-shaped merger remnants opposite to each
  other. Interestingly, we detect strong shocks on to cosmic sheets
  around the haloes at the top and at the bottom.}
\label{fig:halosz2}
\end{figure*}

\begin{figure*}
\centering
\includegraphics{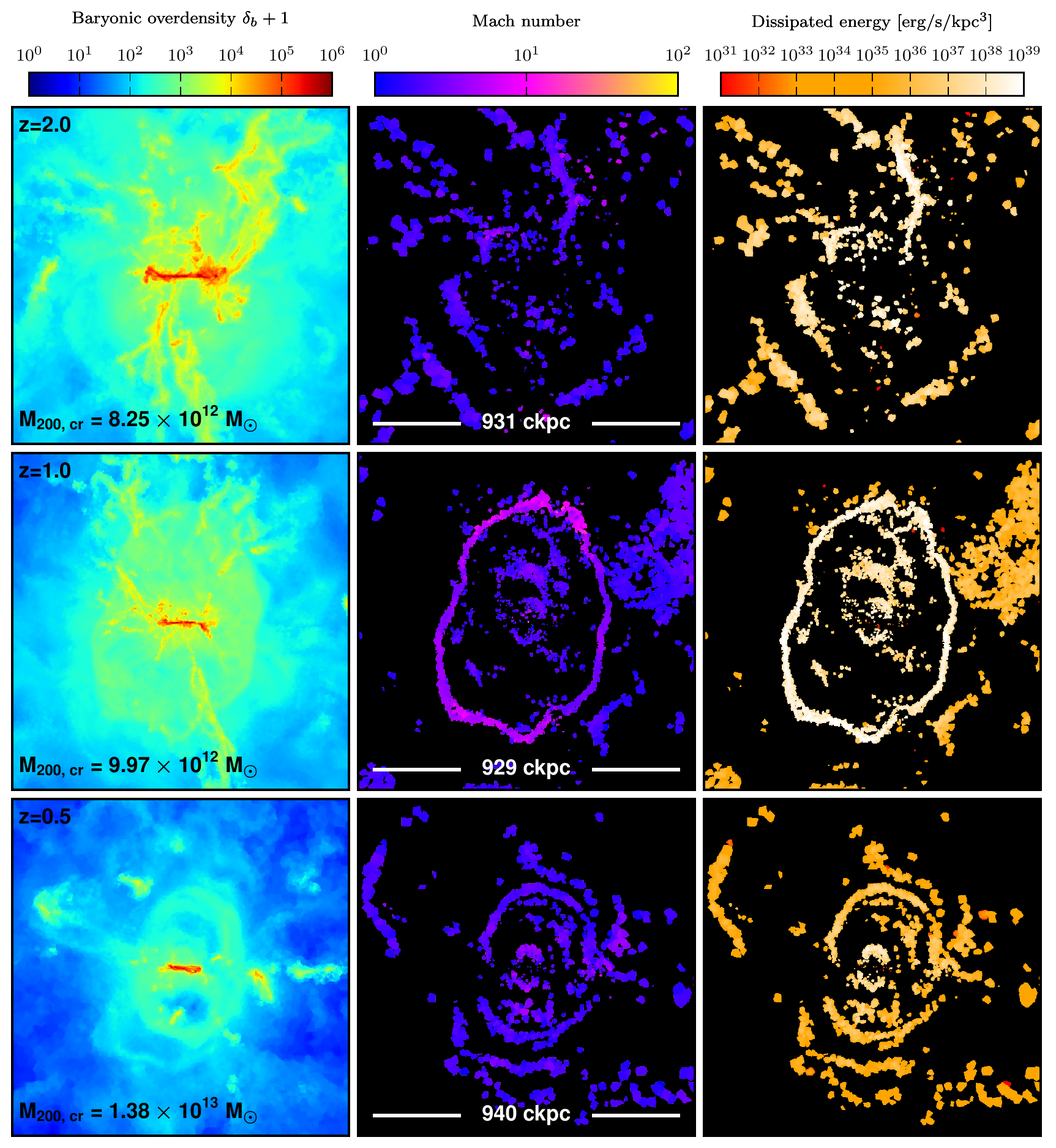}
\caption{Zoom on to disc galaxies at three different redshifts. The
  projections show a square with a size of $1.5\times \rvir$ per side
  and have a depth of $0.08\times \rvir$.  Different morphologies can
  be seen with, from top to bottom, complex inflow and outflow
  patterns, recent BH activity, and supersonic stellar
  winds. Also on these scales the most energetic shocks are provided
  by AGN.}
\label{fig:galaxies}
\end{figure*}

\begin{figure*}
\centering
\includegraphics{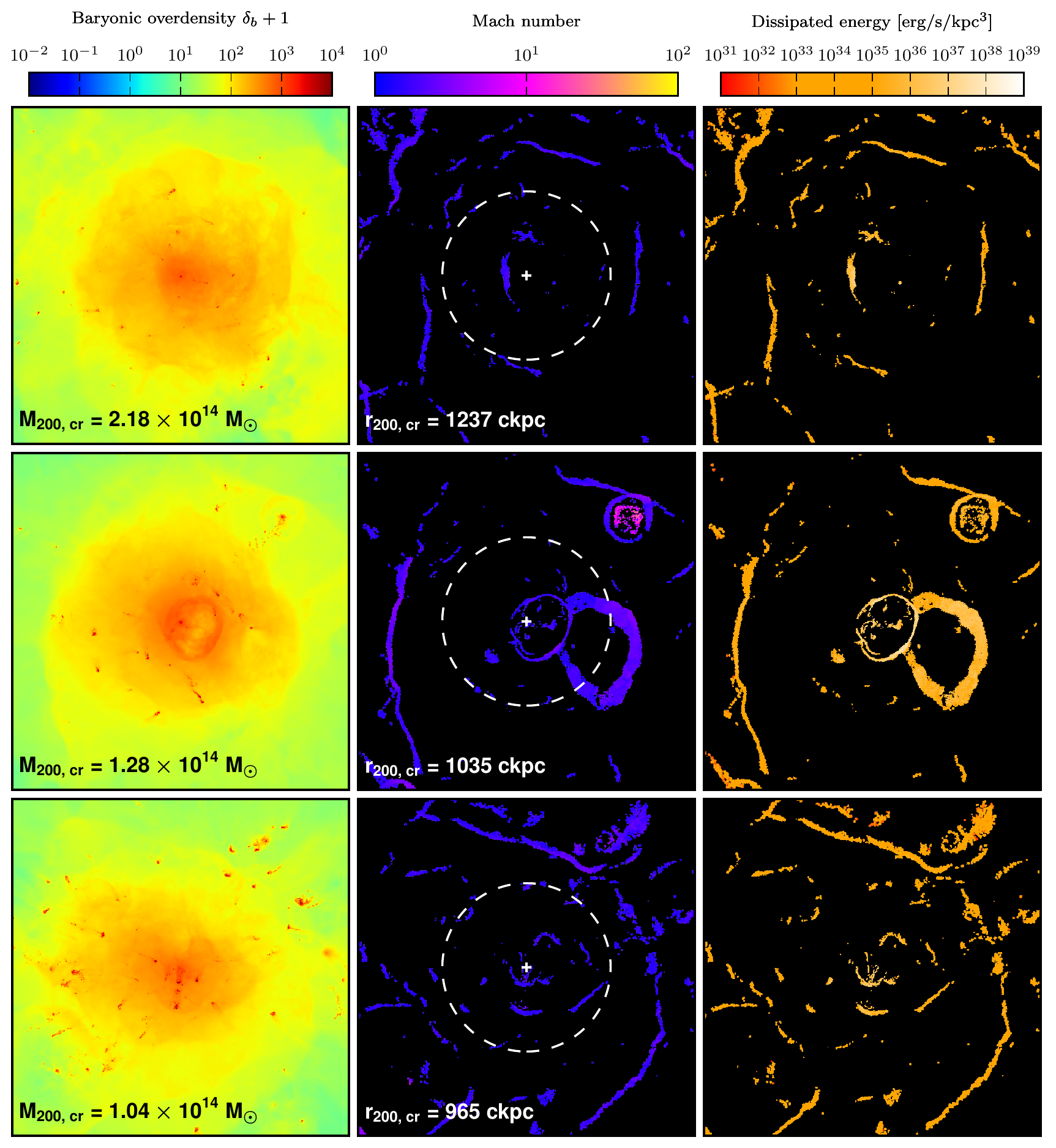}
\caption{Projections of some of the most massive haloes at $z=0$, with
  depths of $0.5 \times \rvir$ and $0.1 \times \rvir$ for the
  overdensity and the shock-related quantities, respectively. Due to
  BH feedback the gas of these haloes has been substantially
  diluted, and most of the present shocks are remnants of this
  process.  However, as can be seen in the middle panels, new BH
  feedback shocks are still created at the present epoch.  In the
  halo at the bottom the accretion of numerous substructures can be
  seen, giving rise to bow-shaped shocks.  }
\label{fig:halosz0}
\end{figure*}

Interestingly, while the filament crosses the outer shock in the vertical
direction, there are also inner accretion shocks in perpendicular
directions, forming as a whole an approximately spherical inner
accretion shock.  This phenomenon can be understood by investigating
the temperature map of the halo, which is shown in the right-hand
panel of Fig.~\ref{fig:velocity_zoom}.  The grey contour line at
$T=3\times10^4\,\mathrm{K}$ corresponds roughly to the pre-shock
temperature of the outer accretion shock, and the black contour line
at $T=1.5\times10^6\,\mathrm{K}$ highlights the inner temperature
structure of the halo. It can be seen that the cold stream from the
filament does not only create a bow shock, it also flows around it and
mixes with gas shock-heated at the outer accretion shock. In this way,
the temperature and sound speed of the latter is decreased such that
it gets shocked a second time in order to reach the virial
temperature.  The black contour line also indicates that the left part
of the inner accretion shock forms mainly due to a cooling stream from
the top; as can be seen in Fig.~\ref{fig:halosz4} there is a
counterpart of the highlighted filament.

This discussion raises the question whether the interpreted shock
surfaces change with simulation resolution, and to which degree they
are modified by radiative physics. We show in
Appendix~\ref{sec:resolution_study} the halo of
Fig.~\ref{fig:velocity_zoom} for different resolutions and find that a
very similar inner accretion shock is obtained in the Illustris-2
run. Moreover, it is also present in the non-radiative simulation
Illustris-NR-2, although in this case the inner shock is less
spherical and the shape is dominated by the two bow shocks in the
directions of the filaments.

It would be instructive to infer the abundance of the double accretion
shocks, ideally in an automated way. However, this proves to be
difficult, since a very similar signature in the differential shock
surface distribution is created by ellipsoidal accretion shocks, which
peak at the radial positions of the semi-principal axes
\citep[see][Fig.~6, Halo 8]{SCHAAL_2015}.  Moreover, the gas
resolution in smaller mass haloes is too low at $z=4$ in Illustris-1
for this kind of analysis.  Nevertheless, it would be interesting in
future simulations to shed light on the questions if, when, and to
which extent gas inside virialized structures of different masses got
shock-heated. This could be accomplished in zoom simulations by
running the shock finder on the fly and recording a shock history for
every gas parcel, a natural extension of a temperature history based 
analysis by means of Monte Carlo tracer particles
\citep{GENEL_2013, NELSON_2013}.

\subsection{Galaxy and galaxy cluster shocks}

In Fig.~\ref{fig:halosz2}, we show the shock environment around some of
the biggest haloes in the simulation at $z=2$. We have chosen these
specific haloes in order to showcase the different occurring shock
morphologies, and they are therefore not necessarily representative
for a typical system.  The cluster at the top accretes streams of
substructures which induce a complex network of low Mach number shocks
in the cluster interior. Due to the high central densities they are
very effective in thermalising kinetic energy, as can be seen in the
top panel on the right-hand side.  Moreover, these merger shocks in
combination with the accretion shocks at the outskirts are thought to
be one of the main drivers of intracluster turbulence
\citep[e.g.][]{DOLAG_2005, VAZZA_2009b, MINIATI_2014, MINIATI_2015}.

Merging clusters often emit unpolarized diffuse synchrotron emission
over large scales, observed as giant radio haloes \citep{FERRARI_2008, 
FERETTI_2012}. These observations indicate the presence of magnetic
fields and relativistic electrons. The latter can acquire their energy
via reacceleration due to magnetohydrodynamic turbulence, or have a hadronic origin
through creation in the decay chain of accelerated protons colliding
with thermal protons (\citealt{ENSSLIN_2011}; for a recent review see
\citealt{BRUNETTI_2014}, and references therein). 
On the other hand, directly accelerating
electrons from the thermal pool via second-order Fermi 
acceleration proves to be difficult from a
theoretical point of view, 
since the required acceleration time is long.
\mybf{Recent limits on the $\gamma$-ray emission of clusters suggest a cosmic ray 
pressure of around 1\%--2\% of the cluster thermal pressure, 
depending on the assumptions \citep{ARLEN_2012, FERMI_2014, AHNEN_2016}.}

The galaxy cluster in the middle panel recently launched an AGN
feedback shock, in the form of a thermal pressure dominated
Sedov--Taylor like blast wave. The Mach numbers across the shock
surface are fairly high, considering also the fact that it has not yet
escaped the hot halo atmosphere. Moreover, this shock is highly
energetic at its current location. We have measured the total energy
dissipation inside the box $[4\times\rvir]^3$ and find that it is
around two times higher compared to the dissipation rate in the cluster
at the top, and around nine times higher than the dissipation of the
cluster shown at the bottom of the figure. For BHs remaining 
predominantly in the radio mode, these kinds of feedback shocks are
launched relatively frequently.  We hence infer that these blast waves
make an important contribution to the high value of the energy
dissipation rate in haloes between $z=3.5$ and $z=2$.

The bottom panels show a galaxy cluster with corrugated shock
surfaces. The surfaces are typically located where the ram pressure of
the inflow matches the thermal pressure of the cluster. In this way,
for homogeneous inflows, the shock locations trace equipotential lines
of the cluster.  In the displayed halo, three major filamentary
streams are present, and the accretion shock shows a concavity in
their direction.  Interestingly, the filaments at the upper left span
a cosmic sheet, and we detect high Mach number accretion shocks at
this location. A similar morphology can be seen for the halo in the
top panels.  Consequently, the large area of Zeldovich pancakes can
provide a very interesting site for particle acceleration near
clusters.  In the central region of the galaxy cluster, merger
remnants with positions opposite to each other are visible; we assume
that this halo underwent a major merger event very recently.

In the local Universe, synchrotron radiation originating from merger
remnants can be observed in the form of radio relics. Considering the
strength and dissipation rate of the AGN feedback shock, we argue that
this shock should also be observable.  Up to now, most of the radio
haloes and relics are observed at redshifts $z\lesssim 0.5$
\citep{FERETTI_2012}, however, this may largely be due to present
sensitivity limits.  But as we show in more detail below, in Illustris
AGN feedback shocks are also created for lower redshifts, up to the
present epoch. Since the shocks appear strong enough to create strong
synchrotron emission, the lack of clear observational detections of
such signals indicates that the radio mode feedback is likely too
strong in Illustris-1. This argument shows that shock properties could
be used in the future to constrain or validate the modelling and
parameter space of feedback models used in state-of-the-art
cosmological simulations.
\mybf{In particular the modelling of radio and gamma-ray emission
represents a promising link between simulations and observations
\citep{PFROMMER_2008a, PINZKE_2010, VAZZA_2013, PINZKE_2015, HONG_2015}.}

The Illustris simulation contains around 40000 well-resolved
galaxies with an encouragingly realistic mix of morphologies and kinematic properties
\citep{VOGELSBERGER_2014,VOGELSBERGER_2014b, GENEL_2014, SNYDER_2015}.
In Fig.~\ref{fig:galaxies}, we show zoomed in view of three late-type
systems with masses of around $10^{13}\,\msol$ at, from top to bottom,
redshifts $z=2$, $z=1$, and $z=0.5$,
in order to investigate shock signatures around such systems. The
displayed images have a length of $1.5\times \rvir$ on a side,
i.e. the virial radius corresponds roughly to the distance between the
corners and the centres of the images.

The spiral galaxy at $z=2$ interacts with the circumgalactic medium
(CGM) via a complex pattern of inflows and outflows, resulting in
several shocks inside the hot halo.  However, this example
demonstrates the limitation of our shock finder in combination with
Illustris.  Poorly resolved gradients in the outskirts of the galaxy
result in patches of shocks, rather than in well-defined shock
surfaces. Moreover, since we filtered out detections of shocks near
star-forming regions that are governed by the subgrid model, it is not
possible to see shocks within the galaxy, for example around spiral
arms.

Recent BH activity is apparent for the galaxy at $z=1$. Given
the shock morphology we suspect that in this case the feedback
originates from the quasar mode. The shape of the shock is less
spherical compared to the blast waves created by the radio-mode
feedback, and moreover, it has apparently been launched very
centrally. This is unlikely to happen in the case of the radio mode
feedback, for which a random injection within a sphere around the
BH is adopted. Nevertheless, the energy dissipation at this
shock is exceedingly high, indicating that also on galactic scales
BH feedback drives the most energetic shocks. This is
especially expected when the BH reaches the end stages of a
phase of exponential growth (during which it shines as a quasar) and
shuts down its own growth through energy feedback
\citep{DiMatteo2005}.

Interestingly, the galaxy at $z=0.5$ launches supersonic galactic
winds, and we are able to detect the associated shocks.  These winds
are dominated by kinetic energy and become particularly effective when
they are strongly collimated in disc-dominated systems.  The role of
SN driven winds in the Illustris galaxy formation model
\citep{VOGELSBERGER_2013} is to slow down star formation, 
suppress cosmological gas inflow \citep{NELSON_2015c}, and
enrich the CGM with metals \citep{BIRD_2014, MARINACCI_2014, SURESH_2015a, SURESH_2015b}.
As the Mach number and energy dissipation fields reveal, the heating through the winds
can be effectively mediated by shocks. We note that the morphology of
galactic wind shocks can be clearly distinguished from other feedback
shocks through the typical consecutive arcs they produce well inside
the virial radius.

Finally, in Fig.~\ref{fig:halosz0}, we show projections of three of
the most massive systems in Illustris-1 at $z=0$.  These cluster-sized
haloes underwent violent BH feedback processes for much of
their cosmic evolution, such that their gas is heavily diluted at
$z=0$.  This also manifests itself in an ambient WHIM gas phase that
is spatially very extended.  The shock morphologies in these
environments are dominated by a small number of very fine surfaces,
indicating that these systems are less dynamic and more relaxed
compared to the massive haloes at higher redshifts.  As can be seen in
Fig.~\ref{fig:energy_comparison}, the energy release rate of BHs 
decreases substantially for $z<0.5$, but the shocks found around
the second halo demonstrate that new strong feedback shock waves are
still created at these times. 

\begin{figure*}
\centering
\includegraphics{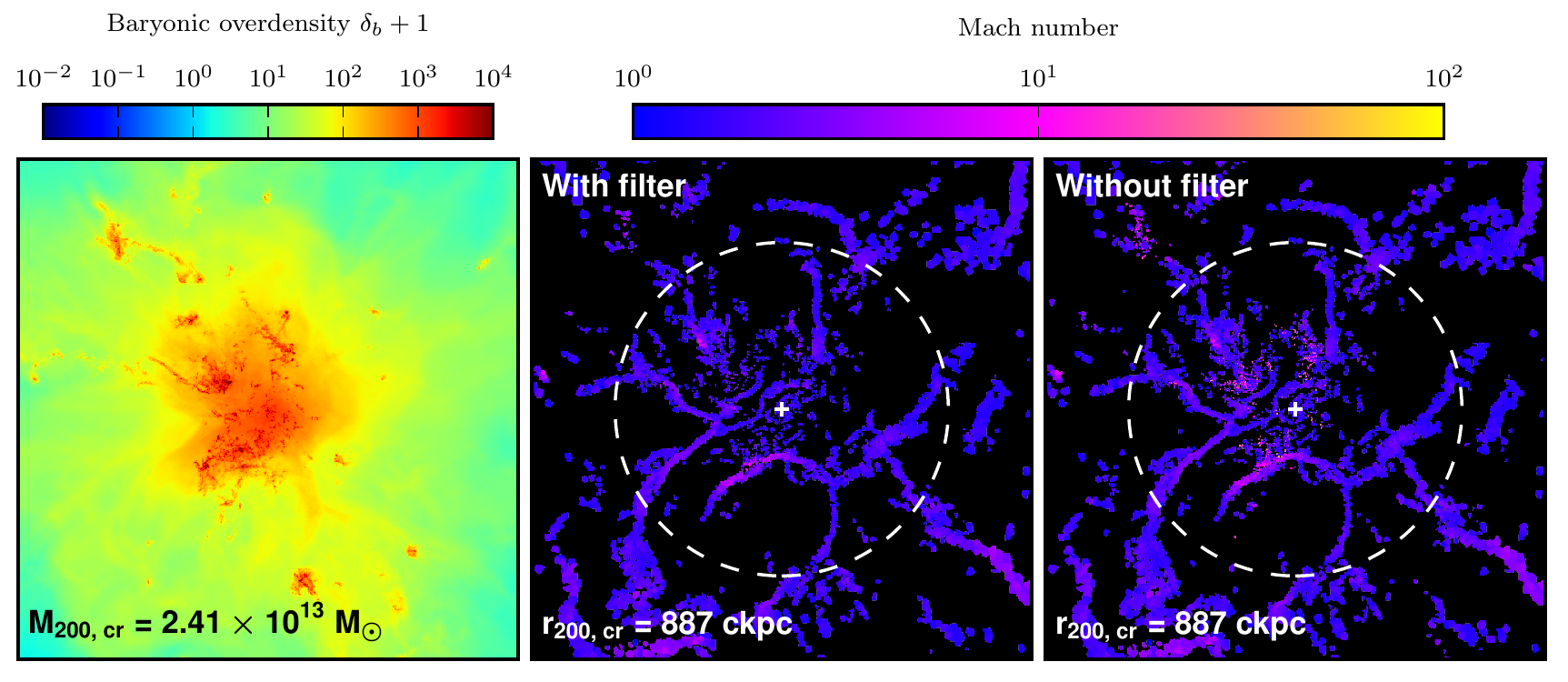}
\caption{Demonstration of the impact of filtering against spurious
  shock detections due to non-hydrodynamic physics.  The middle panel
  shows the shock environment around a halo at $z=2$ obtained with our
  improved shock finding methodology.  In the right-hand side panel
  the result without filtering star-forming gas and inconsistent jumps
  can be seen.  The difference seems to be very subtle, however, the
  spurious detections dominate the total energy dissipation over a
  wide range of redshifts, as shown in Fig.~\ref{fig:variations}.  It
  can therefore be very important to further improve shock finder
  implementations developed for non-radiative simulations before
  applying them to full physics runs.  }
\label{fig:filter_comparison}
\end{figure*}

\begin{figure*}
\centering
\includegraphics{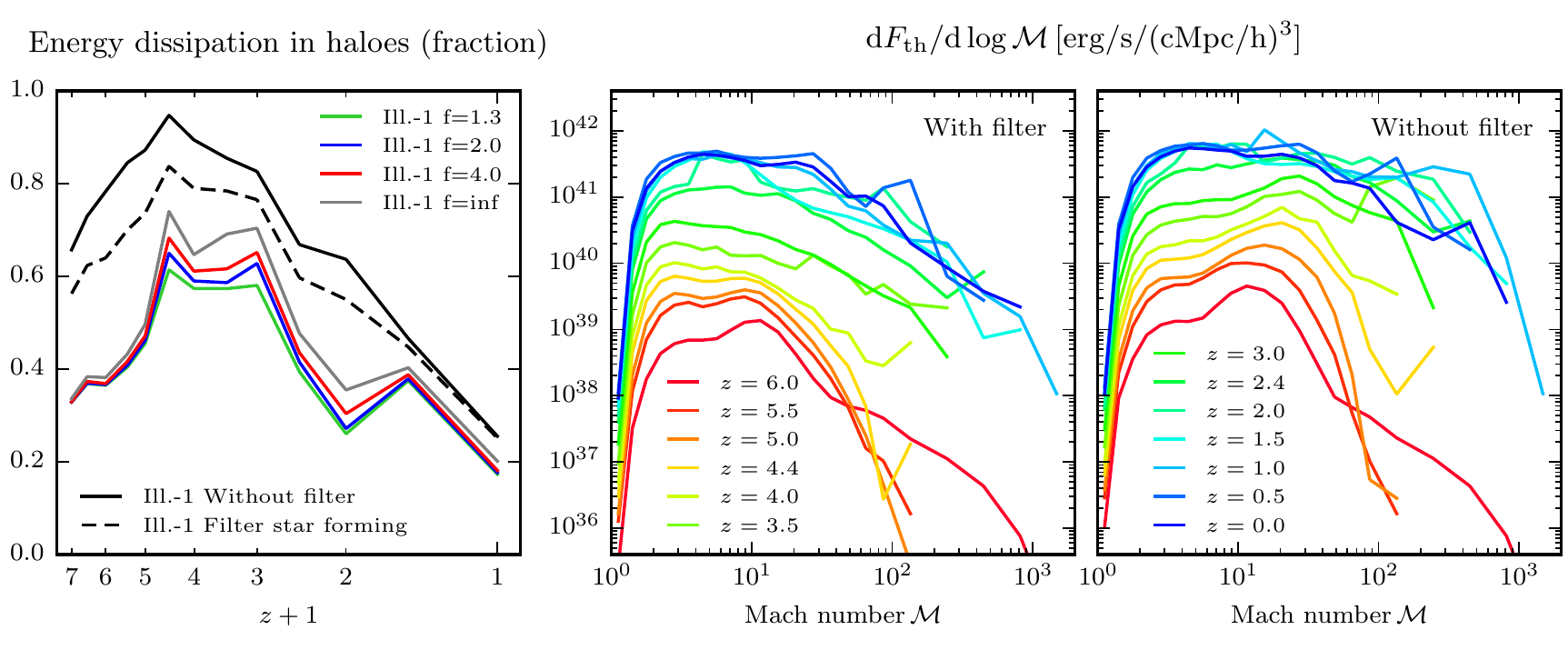}
\caption{{\em Left-hand panel:} contribution of shocks in haloes
  ($\delta_\mathrm{b}\ge100$) to the total dissipation rate for
  different tolerance values $f$, as well as for the case when no
  filter, or only the filter for the star-forming region is used. The
  result does not change significantly for $f\in[1.3,4]$, and hence
  our choice of $f=2$ is very robust. This furthermore indicates that
  unwanted spurious detections are highly inconsistent with the full
  set of Rankine--Hugoniot jump conditions.  {\em Middle and right-hand
    panels:} energy dissipation at shocks as a function of Mach number
  with and without the improved shock finding methodology,
  respectively.  Especially at high redshifts, the spurious detections
  enhance the measured energy dissipation by a factor of several.}
\label{fig:variations}
\end{figure*}

\section{Methodology variations}
\label{sec:methodology_variations}

Our shock-finding algorithm can potentially report spurious detections
of shocks due to large but poorly resolved gradients in regimes which
are dominated by gravitational forces, or where the equations of ideal
hydrodynamics are modified by an imposed equation of state.  This can
for example be the case if gas enters gravitational free fall after
rapid cooling, or for gas around the effective equation of state used
to model the star-forming medium.  The associated jumps in the
hydrodynamic variables should not be confused with shocks. We have
therefore introduced in Section~\ref{sec:shock_finding} the filtering
parameter $f$, which controls the amount of tolerance we allow for
inconsistencies between the jumps inferred from different hydrodynamic
quantities. In this section, we briefly test the dependence of our
results on this nuisance parameter of our shock detection scheme.

In Fig.~\ref{fig:filter_comparison}, we compare the shock environment
around one of the biggest haloes at $z=2$ when no shock filtering is
applied (right-most panel) to the case when active filtering 
with our standard settings is applied (middle panel).  Without the
filter, there is a spray of loosely connected shocked cells with high
Mach numbers whose locations coincide with the highly overdense gas of
star-forming regions. A large fraction of these shocks is likely
spurious and arises because our shock detection algorithm is misled by
the transition to a non-ideal gas equation of state at the star
formation threshold.

\mybf{If the filter is active, cells which are potentially part of the pressurized
ISM, or may have been recently, are excluded from the shock
finding procedure. Moreover, we require for all detections that the 
jump directions of the hydrodynamic quantities are consistent
with the temperature gradient, and for potential shocks having pre-shock 
densities $\delta_\mathrm{b}>1000$ we explicitly verify different
Rankine--Hugoniot jump conditions. As outlined in 
Section~\ref{sec:shock_finding}, only those detections are kept which 
have consistent Mach numbers within a tolerance $f$, inferred by the 
temperature, pressure, and density jump conditions.
For this work we adopt the value $f=2$, which, as can be seen in the middle panel, 
filters most of the unwanted detections. 
Spurious detections are mainly present in high overdensity 
regions, and we find that at $z=0$ around 12\% of all potential shocks
are filtered by our algorithm due to inconsistent jump directions or Mach numbers.}

While in the Mach number projections the spurious detections are only
visible as a subtle feature, they can nevertheless significantly
change the shock energy statistics due to their high pre-shock
densities. In Fig.~\ref{fig:variations}, we compare the energy
dissipation at shocks as a function of Mach number with and without
the filter. In the middle panel, we give the results for our standard
filter configuration as a reference; this plot is the same as in
Fig.~\ref{fig:surface_energy}. The right-hand side panel shows results
obtained if no filter is used.  Especially for high redshifts and Mach
numbers $\machnum > 10$, the measured energy dissipation is greatly
enhanced in this case.  The total dissipation (integral of the curve)
without applied filter is higher by factors of $2.9$, $8.5$, and
$1.4$ at redshifts $z=6$, $z=3.5$, and $z=0$, respectively.

In the left-hand panel of Fig.~\ref{fig:variations}, we investigate
the robustness of our filter with respect to the introduced parameter
$f$.  Shown is the contribution of haloes to the total dissipation
rate for different filter configurations. The black curve is obtained
if no filter is used at all. For the black dashed curve, we filter gas
around star-forming regions, but we do not perform consistency checks.
The grey curve shows the result obtained if the star-forming region is
filtered and in addition the consistency of the jump directions with
the temperature gradient is demanded.  Finally, the coloured curves
indicate results with the full filter being active, but for different
tolerance values $f$.  Even though the parameter range $f\in[1.3,4]$
is rather large, there are only changes at the 10\% level in the
results. This means that the spurious detections are strongly
violating the consistency relations of the Rankine--Hugoniot jump
conditions, allowing them to be reliably filtered with our approach
and a parameter value of $f=2$. Also note that it is not
sufficient to only remove detections around star-forming regions. The
consistency condition is also required to obtain fully reliable
results.

An important lesson from our investigation is that shock finders
designed and tested with non-radiative simulation are unlikely to work
reliably for full physics runs unless they are refined
further. Applying them blindly risks that systematic biases in
measures like the energy dissipation statistics are introduced. The
severity of these biases will depend on the radiative and feedback
physics realized in the simulation code, as well as on the particular
shock finding technique that is used. \mybf{For example, while
\citet{KANG_2007} and \citet{HONG_2014} find only minor effects in runs including cooling
and star formation}, the energy dissipation tail at high Mach numbers
reported in \citet{PFROMMER_2007} might partially originate from
spurious shocks around the adopted subgrid star formation model.

\section{Discussion and summary}
\label{sec:summary}

Cosmological hydrodynamical simulations are now able to successfully
describe the co-evolution of dark matter and baryons in remarkable
detail. As the modelling of the involved physical processes is
refined, it is important to explore novel ways of accessing the
information content provided by such simulations. Thus far, the
analysis of properties of the diffuse gas has often been restricted to
considerations of the most basic quantities, such as (spherically
averaged) density and temperature values. However, gas dynamics has
much more to offer.  In this study, we have for the first time
analysed the shock statistics in a state-of-the-art high-resolution
simulation of galaxy formation, the Illustris project.

We have characterized the shocks in the Illustris universe by
post-processing the simulations with {\small AREPO}'s shock finding
algorithm \citep{SCHAAL_2015}. {\small AREPO} accurately solves the
Euler equations of ideal hydrodynamics with a finite volume method on
a moving-mesh and is able to capture shocks precisely. In
combination with our shock finder we can reliably detect and
characterize the hydrodynamic shocks and associated quantities.

From a theoretical point of view, shocks are of fundamental
interest. They arise from the non-linearity of the hydrodynamic
equations and are therefore a manifestation of one of the key
characteristics of the governing equations.  In order to understand
the gas dynamics in a complex system, much can be learned by merely
investigating the locations and strengths of shocks. Moreover, shock
quantities provide a reduced dataset of the gas in hydrodynamic
cosmological simulations, which can be used to simplify the analysis
of large output files\footnote{In the case of Illustris-1, a snapshot
consists of around $1.2\times 10^{10}$ resolution elements and has
a size of about 2.5 terabytes.}.

In this work we have focused on analysing shocks in the full physics
Illustris-1 simulation. In order to put the results for the most
important statistics into context, we also compared with results
obtained for the non-radiative run llustris-NR-2.  The differences in
shock statistics and shock morphologies between the two kinds of
simulations arise due to physics beyond ideal gas dynamics in
Illustris-1.  Most importantly, this simulation accounts for cosmic
reionization, radiative cooling and heating of gas, star formation and
BH growth, as well as their associated feedback processes.  In
what follows, we summarize our most important findings.

\begin{itemize}
\item {\bf Improvement of the shock finding methodology.} In the full
  physics runs, we have discovered spurious shock detections in
  star-forming regions and at the edges of cold, poorly resolved
  self-gravitating clumps. They originate from modifications of the
  gas state by the subgrid ISM model and cooling physics,
  respectively. We have shown that these spurious detections severely
  change the energy dissipation statistics if the problem is just
  ignored. We have thus introduced improvements to our shock
  finding routines that were not part of the method we developed
  originally for non-radiative runs.
  First of all, we excise star-forming regions in our shock finding
  algorithm. Secondly, we ignore detections for which the hydrodynamic
  jumps are inconsistent with the shock direction given by the
  temperature gradient, and last but not least, we demand that shocks
  in overdense regions consistently fulfil the Rankine--Hugoniot jump
  conditions for several different hydrodynamic quantities. We have
  demonstrated that these conservative filters are necessary to
  reliably find and characterize hydrodynamic shocks in the Illustris
  simulation. We anticipate similar problems for other shock finders
  developed for non-radiative runs, and hence caution against applying
  them in unmodified form to full physics runs.

\item {\bf Impact of reionization.}  We find in Illustris-1 a
  significant increase in the measured shock surface area as well as in
  the total energy dissipation between redshifts $z=6$ and $z=5.5$.
  This coincides with reionization, which is modelled by a uniform UV
  background following \citet{FAUCHER_2009} and happens in the
  simulation almost instantaneously at $z\lesssim 6$.  We interpret
  this finding as the signature of a temporary population of weak
  shocks associated with the dynamical response of the gas to the
  sudden heating during reionization, which for example lets gas
  stream off the shallow potential wells of very low-mass dark matter
  haloes. Quantitatively, this result certainly depends on the details
  of the reionization transition.

\item {\bf Shock surface area.} Until $z\approx 1$ by far the
  largest shock surface area can be found between voids and the cosmic web.  
  The cumulative shock surface statistic is therefore dominated by the number and
  temperature of voids. During structure formation, small voids merge
  to form larger ones, and the number of low Mach number shocks
  decreases. At the same time, voids are coupled to the Hubble flow
  and cool adiabatically, resulting in an increase in the occurrence of
  high Mach number shocks.  Moreover, we find that the number of
  shocks with $\machnum\approx 4$ stays roughly constant with time.
  For $z<1$ a significant number of shocks can also be found in the
  WHIM, with a maximum contribution of around $34\%$ at $z=0$.  After
  reionization, the total shock surface area decreases monotonically,
  nevertheless, at the present epoch we find a total shock surface 
  area in Illustris-1 which is around 1.4 times larger
  compared to the corresponding non-radiative run.

\item {\bf Energy dissipation.}  By measuring the energy dissipation
  rate at shocks the impact on the thermal history of the gas can be
  inferred. In non-radiative runs, the overall energy dissipation grows
  with cosmic time, and when its Mach number dependence is considered
  a bimodality in the energy statistics can be seen; internal low Mach
  number shocks dissipate most of the energy, whereas external shocks
  have high Mach numbers but are much less energetic \citep{RYU_2003}.
  Remarkably, we find for Illustris-1 that shocks with Mach numbers
  smaller and larger than $\machnum\approx 10$ contribute about
  equally to the overall dissipation, and no bimodality is observed.  For this
  full physics simulation, we attribute the large population of high
  Mach number shocks to feedback processes.  

  Between $z=3.5$ and $z=2$, most of the energy dissipation occurs
  within haloes, where cooling processes create structures with large
  density contrasts. It is also around this time that the radio mode
  feedback of BHs first kicks in, which is associated with the
  low accretion rate state of the BHs. We show that this type
  of feedback drives energetic high Mach number shocks and creates an
  extended WHIM gas phase. We measure a total energy dissipation in
  Illustris-1 which is higher by a factor of around 8 for $z<2$
  compared to the corresponding non-radiative run. It is plausible
  that a large fraction of this additional dissipation originates from
  the BH radio mode feedback.

  As a result of the redistribution of baryons at low redshifts to the
  WHIM, this phase gains importance as a site for energetic shocks.
  At $z=0$ around $80\%$ of the dissipation occurs at shocks in the
  WHIM and in the hot diffuse phase ($T>10^7$ and $\delta_b<100$).
  This is in strong contrast to the corresponding non-radiative
  simulation, in which about $60\%$ of the thermalization happens
  inside haloes at the present epoch. At $z=0$ the total dissipation
  rate in shocks in Illustris-1 amounts to
  $1.7\times
  10^{41}\,\mathrm{erg}\,\mathrm{s}^{-1}\,\mathrm{Mpc}^{-3}$.
  Moreover, except for the cold diffuse gas phase, all environments
  have similar and approximately constant specific dissipation rates
  for $z<3$.  This value is approximately
  $10^{-1}\,\mathrm{erg}\,\mathrm{g}^{-1}\mathrm{s}^{-1}$.

\item {\bf Shock morphologies.}  We have investigated accretion shocks
  of haloes with mass above $10^{12}\,\msol$ at $z=4$ when feedback
  processes influencing large scales are still largely absent. As
  expected for haloes in this mass range, we find accretion shocks
  which are typically located outside of the virial radius. However,
  we see filaments which can penetrate the accretion shock, and their
  gas streams enter the halo unshocked. Interestingly, we discover
  that the formation of a spherical accretion shock further inside the
  halo is still possible in this case. Our finding of an accretion
  mode involving both an inner and an outer accretion shock is
  consistent with recent work by \citet{NELSON_2015}, who analysed the
  thermal gas structure of $10^{12}\,\msol$ haloes at $z=2$ in zoom simulations.

  We have showcased different shock morphologies of galaxies and
  galaxy clusters at $z\leq2$. Those include complex networks of weak
  shocks in cluster interiors, merger remnants, strong BH
  feedback shocks, shocks on to cosmic sheets, as well as supersonic
  stellar winds and complex flows around galaxies.  By visually
  comparing the properties of an AGN feedback shock with a merger
  remnant shock, we argue that the radio mode feedback is too strong
  in Illustris-1. This conclusion is consistent with the findings of
  \citet{GENEL_2014} and \citet{HAIDER_2015}, who report a baryon
  fraction inside clusters which is too small at the present epoch
  and also attribute this finding to the radio mode feedback channel.
  This also shows that the properties of shocks can be used in future
  state-of-the-art cosmological simulations for constraining the
  modelling and parameter space of feedback implementations.

\end{itemize}

These results underline the importance of feedback processes for shock
statistics, and the large impact they have on the shocks that are
present. As we expect the strongest shocks to be efficient
accelerators that produce non-thermal particle distributions, shock
statistics combined with upper limits on cosmic ray densities could be
developed into a powerful constraint for viable feedback
models. Future simulation models will also try to take the cosmic ray
particle populations self-consistently into account.
\mybf{In this case further refinements of the shock 
detection methodology are needed in order to cope with 
mixtures of ordinary gas and cosmic rays \citep{PFROMMER_2016}.}

\section*{Acknowledgements}

It is a pleasure to thank Andreas Bauer, Rainer Weinberger, Christian
Arnold, Christine Simpson, Robert Grand, Martin Sparre, Federico
Marinacci, Rahul Kannan, Lorenzo Sironi, and Blakesley Burkhart for
useful comments and insightful discussions.  
\mybf{Moreover, the authors would like to thank the referee
for a constructive report which helped improving this paper.}
KS and VS acknowledge
support through subproject EXAMAG of the Priority Programme 1648
SPPEXA of the German Science Foundation, and the European Research
Council through ERC-StG grant EXAGAL-308037.
CP acknowledges support through the ERC-CoG grant CRAGSMAN-646955.
SG acknowledges support provided by NASA through Hubble Fellowship 
grant HST-HF2-51341.001-A awarded by the STScI, which is operated 
by the Association of Universities for Research in 
Astronomy, Inc., for NASA, under contract NAS5-26555.
DS acknowledges support by the STFC and the ERC Starting 
Grant 638707 ``Black holes and their host 
galaxies: co-evolution across cosmic time''.
LH acknowledges support from NASA grant NNX12AC67G and NSF grant
AST-1312095. KS, VS, RP, and CP like to thank
the Klaus Tschira Foundation, and KS acknowledges support by the IMPRS for
Astronomy and Cosmic Physics at the University of Heidelberg.
\bibliographystyle{mnras} \bibliography{literature}

\appendix
\section{Resolution study}
\label{sec:resolution_study}

\begin{figure*}
\centering
\includegraphics{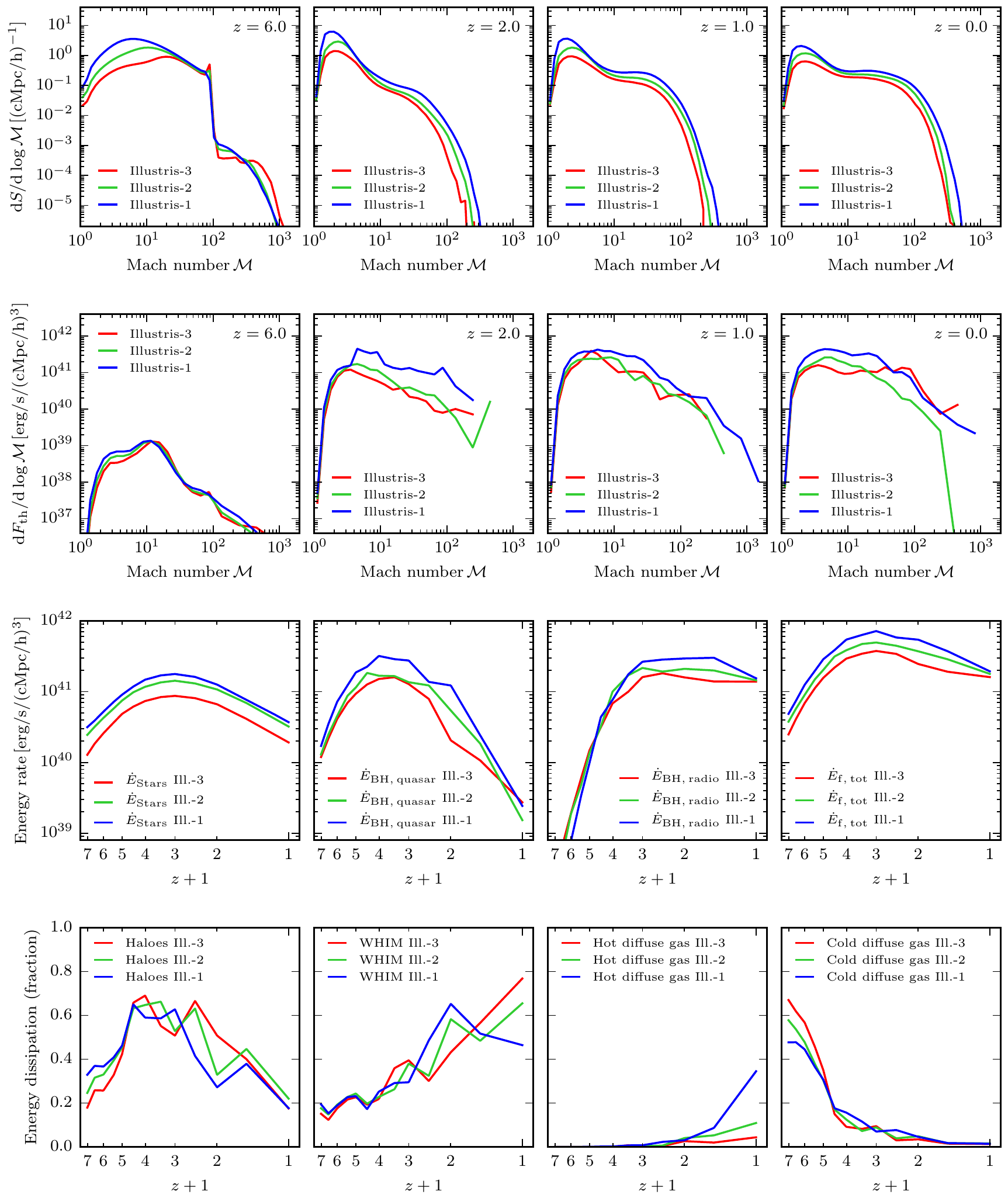}
\caption{Convergence study of several quantities of the full physics
  runs. From top to bottom we show the cumulative shock surface as a function
  of Mach number, the energy dissipation at shocks as a function of
  Mach number, the energy released in different feedback channels as a
  function of redshift, and the corresponding energy dissipation
  within different environments as a function of redshift.  A higher
  BH feedback energy rate is present in Illustris-1 compared
  to the lower resolution runs, and we suspect this to be the main reason
  for the increased energy dissipation rate at shocks for $z\leq 2$.}
\label{fig:convergence}
\end{figure*}

\begin{figure*}
\centering
\includegraphics{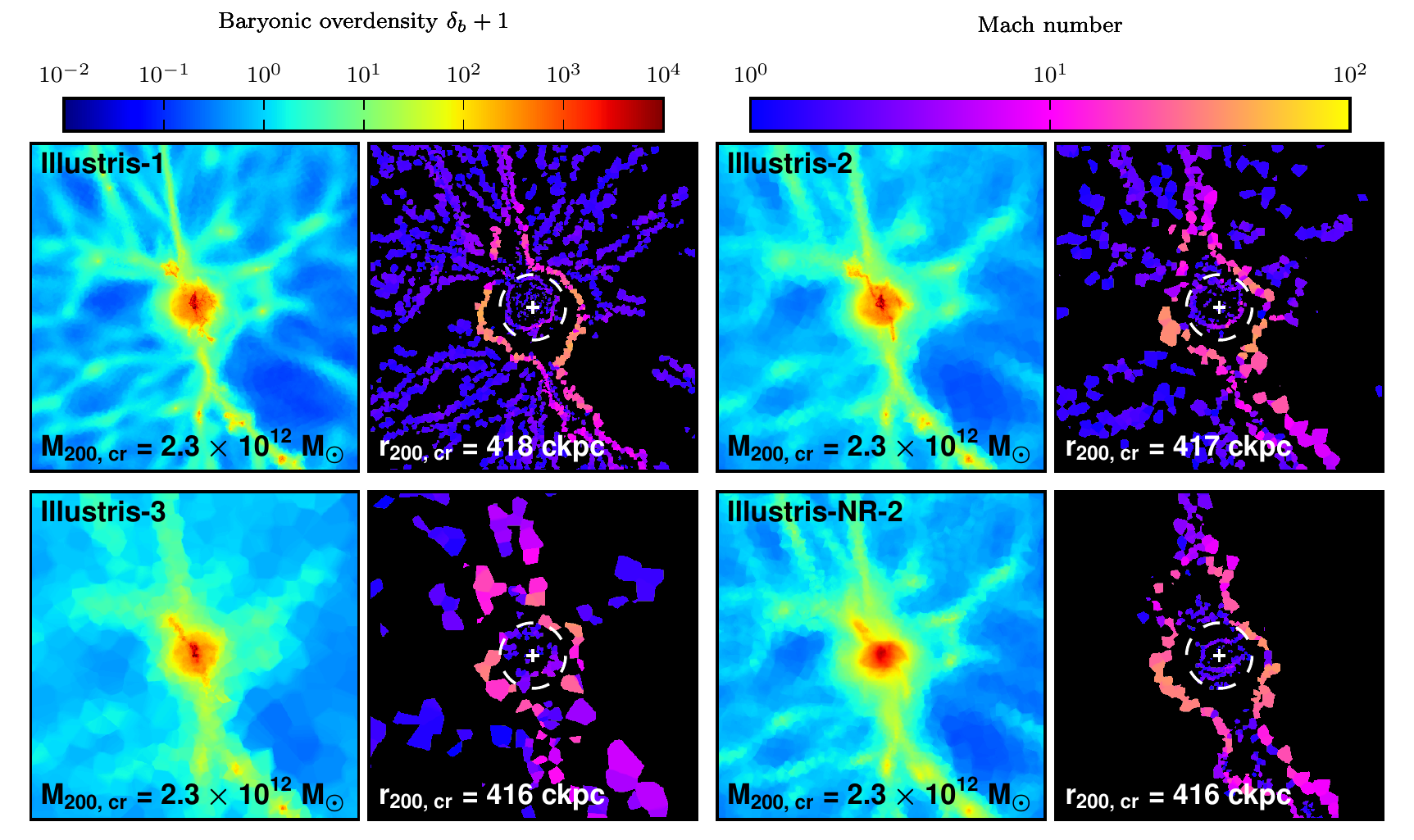}
\caption{Resolution study for the halo with double accretion shock
  shown in Fig.~\ref{fig:velocity_zoom}.  Although less details are
  resolved in the lower resolution runs, the inner accretion shock can
  still be seen.  Moreover, it is also present in the non-radiative
  run Illustris-NR-2, providing evidence that this shock is truly an
  accretion phenomenon.  In the full physics runs the shape of the
  inner shock is roughly spherical, whereas in the non-radiative run
  it consists of connected bow-shaped shocks, suggesting that cooling
  helps to stabilize the inner accretion shock and make them more
  spherical. }
\label{fig:halo_convergence}
\end{figure*}

In Fig.~\ref{fig:convergence}, we study resolution effects related to
shock properties for the full physics runs Illustris-1, Illustris-2,
and Illustris-3.  Illustris-2 and Illustris-3 have an 8 and
64 times lower mass resolution compared to Illustris-1,
respectively, but adopt the same physics model and 
evolve otherwise identical initial conditions. The first row shows
the cumulative shock surface area as a function of Mach number for the different
resolutions. The results across cosmic time are similar, with the
trend that more shocks are found in higher resolution runs. This is
especially true at high redshift. These findings are expected since
finer and therefore more numerous structures are resolved with
increasing resolution.

In the second row we investigate resolution effects on the energy
dissipation at shocks as a function of Mach number. At $z=6$ we find 
very good agreement, indicating that the difference in the cumulative  
shock surface area at this time is due to less energetic external shocks.  
At $z\leq 2$, shocks with $\machnum\gtrsim 5$ dissipate significantly
more energy in the highest resolution run compared to Illustris-2 and
Illustris-3. As can be seen in the left-hand side panel of
Fig.~\ref{fig:energy_comparison}, this results in a total energy
dissipation which is higher by a factor of around 2 at late times.

We suspect that the origin of this discrepancy lies in the feedback
energy released by BHs, for the following reasons.  The energy
output of quasar and radio AGN as a function of redshift is shown in
the central panels of the third row, and for both channels
significantly larger values are reached in the highest resolution run.
Moreover, as can be seen in the row at the bottom, a notable hot
diffuse phase is only created in Illustris-1. \mybf{As we have argued in
Section \ref{subsub:envtime},} this phase originates from
few and highly energetic shocks.  Nevertheless, we obtain overall
reasonably robust results. Given that the simulations themselves have
to produce resolution independent results, as well as the shock
finding algorithm, this achievement is non-trivial.

Fig.~\ref{fig:halo_convergence} shows halo 4 of
Fig.~\ref{fig:halosz4}, which can also be seen in the zoom of
Fig.~\ref{fig:velocity_zoom}, for different simulations.  It can be
seen, as previously discussed, that fewer shocks are found for the
lower resolution runs. Moreover, in the non-radiative simulation
Illustris-NR-2, many shocks on to the cosmic web are erased by the
global temperature floor imposed in post-processing to model
reionization.  In Illustris-3, the resolution is not high enough for
detecting distinct shock surfaces at this scale and redshift ($z=4$).
Nevertheless, the detections at hand point towards the existence of a
double accretion shock.  On the other hand, in Illustris-1 and
Illustris-2, the inner and the outer accretion shock are clearly
visible, and moreover, the measured Mach numbers are similar. We
hence expect that these surfaces are largely converged; running the
simulation with an even higher resolution would result in shocks at
very similar locations and with very similar properties.

The inner accretion shock is also present in the non-radiative run
Illustris-NR-2, and is hence unrelated to feedback
processes. Remarkably, the inner shock is less spherical in this
simulation and consists of two connected bow-shaped shocks pointing in
the direction of the filaments. This observation leads us to suggest
that the sphericity in the full physics runs originates from cooling
physics, as discussed in Section~\ref{sec:highzaccr}.

\bsp	
\label{lastpage}
\end{document}